\documentclass[sigconf, nonacm]{acmart}

\newcommand\vldbdoi{XX.XX/XXX.XX}
\newcommand\vldbpages{XXX-XXX}
\newcommand\vldbvolume{14}
\newcommand\vldbissue{1}
\newcommand\vldbyear{2020}
\newcommand\vldbauthors{\authors}
\newcommand\vldbtitle{\shorttitle} 
\newcommand\vldbavailabilityurl{URL_TO_YOUR_ARTIFACTS}
\newcommand\vldbpagestyle{plain}

\usepackage{mdframed}
\usepackage{enumitem}
\usepackage{pifont}
\usepackage{bm}
\usepackage{graphicx}
\usepackage{svg}
\usepackage{amsmath}
\usepackage{amssymb}
\usepackage{amsfonts}
\usepackage{float}
\usepackage{array}
\usepackage{siunitx}
\usepackage{tabularx}
\usepackage{makecell}
\usepackage{diagbox}
\usepackage{multirow}
\usepackage{xcolor}
\usepackage{booktabs}
\usepackage{adjustbox}
\usepackage{tikz}
\usepackage{xcolor}
\usepackage{tcolorbox}
\usepackage{stfloats}
\usepackage{hyperref}

\newcommand{\centeredItalic}[1]
{
    \begin{center}
        \textit{#1}
    \end{center}
}

\definecolor{Maroon}{RGB}{128, 0, 0} 
\definecolor{MyGreen}{HTML}{A8E6A1}

\newmdenv[
    backgroundcolor=gray!10,
    linewidth=0pt,
    innerleftmargin=3pt,
    innerrightmargin=3pt,
    innertopmargin=3pt,
    innerbottommargin=3pt,
    skipabove=3pt,
    skipbelow=3pt,
    leftmargin=1mm,
    rightmargin=1mm,
    frametitlefont=\bfseries,
    frametitlealignment=\hspace{0pt},
    linecolor=white
]{mytakeaway}

\newmdenv[
    backgroundcolor=green!10,
    linewidth=0pt,
    innerleftmargin=3pt,
    innerrightmargin=3pt,
    innertopmargin=3pt,
    innerbottommargin=3pt,
    skipabove=3pt,
    skipbelow=3pt,
    leftmargin=1mm,
    rightmargin=1mm,
    frametitlefont=\bfseries,
    frametitlealignment=\hspace{0pt},
    linecolor=white
]{myinsight}

\newmdenv[
    backgroundcolor=yellow!10,
    linewidth=0pt,
    innerleftmargin=3pt,
    innerrightmargin=3pt,
    innertopmargin=3pt,
    innerbottommargin=3pt,
    skipabove=3pt,
    skipbelow=3pt,
    leftmargin=1mm,
    rightmargin=1mm,
    frametitlefont=\bfseries,
    frametitlealignment=\hspace{0pt},
    linecolor=white
]{myconclusion}

\begin{document}
\title{HERO: Hint-Based Efficient and Reliable Query Optimizer}

\author{Zinchenko Sergey}

\author{Iazov Sergey}

\begin{abstract}
    We propose a novel model for learned query optimization which provides query hints leading to better execution plans. The model addresses  the three key challenges in learned hint-based query optimization: reliable hint recommendation (ensuring non-degradation of query latency), efficient hint exploration, and fast inference.  We provide an in-depth analysis of existing NN-based approaches to hint-based optimization and experimentally confirm the named challenges for them. Our alternative solution consists of a new  inference schema based on an ensemble of context-aware models and a graph storage for reliable hint suggestion and fast inference, and a budget-controlled training procedure with a local search algorithm that solves the issue of exponential search space exploration. In experiments on standard benchmarks, our model demonstrates optimization capability close to the best achievable with coarse-grained hints. Controlling the degree of parallelism (query dop) in addition to operator-related hints enables our model to achieve 3x latency improvement on JOB benchmark which sets a new standard for optimization. Our model is interpretable and easy to debug, which is particularly important for deployment in production.
\end{abstract}

\maketitle

\pagestyle{\vldbpagestyle}
\begingroup\small\noindent\raggedright\textbf{PVLDB Reference Format:}\\
\vldbauthors. \vldbtitle. PVLDB, \vldbvolume(\vldbissue): \vldbpages, \vldbyear.\\
\href{https://doi.org/\vldbdoi}{doi:\vldbdoi}
\endgroup
\begingroup
\renewcommand\thefootnote{}\footnote{\noindent
This work is licensed under the Creative Commons BY-NC-ND 4.0 International License. Visit \url{https://creativecommons.org/licenses/by-nc-nd/4.0/} to view a copy of this license. For any use beyond those covered by this license, obtain permission by emailing \href{mailto:info@vldb.org}{info@vldb.org}. Copyright is held by the owner/author(s). Publication rights licensed to the VLDB Endowment. \\
\raggedright Proceedings of the VLDB Endowment, Vol. \vldbvolume, No. \vldbissue\ %
ISSN 2150-8097. \\
\href{https://doi.org/\vldbdoi}{doi:\vldbdoi} \\
}\addtocounter{footnote}{-1}\endgroup

\ifdefempty{\vldbavailabilityurl}{}{
\vspace{.3cm}
\begingroup\small\noindent\raggedright\textbf{PVLDB Artifact Availability:}\\
The source code, data, and/or other artifacts have been made available at \url{\vldbavailabilityurl}.
\endgroup
}

\section{Introduction}
\label{chapter:introduction}

Query optimization is one of the key research areas in data management \cite{chaudhuri2011overview}, since the performance of a DBMS directly depends on the quality of query execution plans \cite{chaudhuri1998overview}. As traditional query optimizers often face issues in producing efficient plans \cite{lohman2014query}, recently, quite a number of Machine Learning (ML) solutions have been proposed to address them including learned cost models \cite{siddiqui2020cost}, cardinality estimators \cite{kipf2018learned,negi2020cost,negi2021flow,sun2021learned,marcus2018deep}, and E2E learned optimizers \cite{marcus2019neo,yang2022balsa}. However, all these approaches face the problem of generalization and eventual query performance degradation \cite{ortiz2019empirical} which makes them risky for deployment in production. An alternative hint-based optimization model has been first proposed in \cite{marcus2020bao} which can operate on top of an existing optimizer by guiding it with \textit{hints} potentially leading to better query plans. However as we demonstrate in our paper, the approach of \cite{marcus2020bao} (and its successors) still suffers from the \textit{main two issues}: the lack of guarantees that suggested hints non-degrade query latency and long training (hint exploration) and inference times. The main reason for eventual query degradation is the use of NN-based models with which, as we show in our paper, it is hard to predict reliable hints, while the reason for the long inference is the need to deal with an exponentially large space of hint combinations.

In this paper, we propose a novel \underline{h}int-based \underline{e}fficient and \underline{r}eliable \underline{o}ptimizer (HERO) that addresses the problems of existing solutions. The contribution of our work can be summarized as follows.
\begin{itemize}
    \item We provide a formalization of the problem of reliable hint-based optimization (\S \ref{chapter:problem_statement}).
    \item We describe the architecture of HERO and ideas behind its components (\S \ref{chapter:hero_design}). To overcome long hint exploration and inference times, we propose a parameterized local search algorithm adaptable to varying time budgets. Additionally, we present a more lightweight and reliable alternative to NN models which solves the issue of prediction reliability. Our solution is provided by an ensemble of context-aware models making HERO interpretable, transparent, and easy to debug, which is important for real-world application.  
    \item We provide an in-depth study of existing NN-based models for hint-based optimization (\S \ref{chapter:pipeline}) and present results of a detailed experimental analysis of NN behavior (\S \ref{chapter:justification}) highlighting the core difficulties in achieving reliable generalization.
    \item We provide experimental results demonstrating that HERO offers a safer and faster solution (\S \ref{chapter:experimental_evaluation}) and achie\-ves a higher optimization ratio than the existing NN-based approaches. We also show that HERO components can be combined with NN-models to further boost optimization in scenarios when potential performance gain for some queries is prioritized over eventual degradations.
    \item We provide open-source datasets that we used in our study on the limits of NN-based approaches. These include NN training code,\footnote{\href{https://github.com/handdl/btcnn}{https://github.com/handdl/btcnn}} a reusable platform for rapid evaluation of different query acceleration strategies\footnote{\href{https://github.com/zinchse/hbo_bench}{https://github.com/zinchse/hbo\_bench}}, model weights, and extended experimental results.\footnote{\href{https://github.com/zinchse/hero}{https://github.com/zinchse/hero}} We believe that the platform and datasets can be used as a testbed in the development of novel hint-based optimization tools avoiding the need to run costly query optimization experiments on a DBMS.

\end{itemize}

\noindent \textbf{Method of Experimental Evaluation.} 
For experimental evaluation, we used two IMDb-based benchmarks: JOB benchmark \cite{leis2015good} consisting of 113 queries and its skewed version SQ (sample\_queries from the repository\footnote{\url{https://github.com/learnedsystems/baoforpostgresql}} of \cite{marcus2021bao}) with 40 queries. Additionally, we used TPCH benchmark \cite{tpch}, which consists of 22 queries, with a scale factor of 10. For fast experimentation with tens of thousands of different hint exploration strategies we implemented the following approach: for every query from these benchmarks and for all possible hint combinations, we \underline{\textit{saved}} execution plans and their latencies obtained on openGauss DB \cite{opengauss}. This allowed us to \textit{replace actual query execution with a simple table} \underline{\textit{lookup}}. To ensure the consistency of the collected data, the server was exclusively used during idle periods, with statistics updates disabled, and the database pre-warmed before each query execution. The obtained openly published dataset can be \textit{re-used} to accelerate experimentation with various query exploration strategies in hint-based optimization.

\noindent \textbf{Experimental Setup.} For all our experiments, we used a 128-core Kun\-peng-920 (2.6 GHz) running openGauss RDBMS. The database server was configured with 200 GB of process memory, 100 GB of cstore\_buffers, 80 GB of work\_mem, and 160 GB of shared\_buffers.

\section{Related Work}
\label{chapter:related_work}

One of the first approaches to learned hint-based optimization is \textbf{Bao} \cite{marcus2020bao}, which employs the idea of generating query plans with \textit{all} possible hint combinations and evaluating them independently by using a trained NN. While this allows for potentially achieving optimal performance, in practice, Bao faces challenges with \textit{long training} and \textit{inference} times. Additionally, due to the black-box nature of its NN model, it suffers from poor interpretability and \textit{unreliable} predictions. In \cite{negi2021steering}, some of these problems were solved with a randomized algorithm and heuristics. However, it required a deep integration into the optimizer and still depended on expensive planning. As a result, this solution is only applicable to large-scale workloads, such as Microsoft's SCOPE.

In \textbf{QO-Advisor} \cite{zhang2022deploying} deployed in Microsoft’s production systems, the work of Bao was further extended. To address the risk of performance degradation, the authors incorporated an \textit{A/B testing infrastructure} into the pipeline for additional validation and exploration, while limiting hint exploration to local changes in only a \textit{single} hint. Although this solution mitigates the degradation issue, it also \textit{limits the potential} of hint-based optimization.

In \textbf{AutoSteer} \cite{anneser2023autosteer}, the authors proposed to iteratively explore the hint search space by using a greedy algorithm at the training and inference stages. As we have confirmed in experiments, the greedy algorithm tends to skip exploring many potential states, sometimes overlooking highly effective ones, which prevents it from implementing the full potential of hint-based optimization. Furthermore, due to the sequential nature the greedy algorithm can not be parallelised which leads to a big \textit{planning overhead} at inference and makes the approach impractical.

In \textbf{FastGres} \cite{woltmann2023fastgres} instead of using a single NN the authors proposed to used several context-aware models. This approach allowed them to find a balance between the complexity and reliability. However, the authors did not pay attention to the resulting plan, without which, as we show in this paper, it is impossible to make reliable predictions. For this reason, we do not include FastGres into the list of competitors in our experimental comparison.

\begin{table}
\centering
\resizebox{\columnwidth}{!}{
\begin{tabular}{c c c c c}
\toprule
& \multicolumn{2}{c}{\textbf{Exploration}} & \multicolumn{2}{c}{\textbf{Inference}} \\
\cmidrule(lr){2-3} \cmidrule(lr){4-5}
\textbf{Model} & \textbf{Time $\downarrow$} & \textbf{Quality $\uparrow$} & \textbf{Time $\downarrow$} & \textbf{Reliability $\uparrow$} \\
\midrule
HERO & \textbf{low} & \textbf{high} & \textbf{low} & \textbf{high} \\
Bao \cite{marcus2020bao} & \textcolor{Maroon}{\textbf{high}} & \textbf{high} & \textcolor{Maroon}{\textbf{high}} & \textcolor{Maroon}{\textbf{low}} \\
AutoSteer \cite{anneser2023autosteer} & medium & medium & \textcolor{Maroon}{\textbf{high}} & medium \\
QO-Advisor \cite{zhang2022deploying} & \textbf{low} & \textcolor{Maroon}{\textbf{low}} & \textbf{low} & \textbf{high} \\
\bottomrule
\end{tabular}
}
\caption{Comparison of hint-based optimizers on static workloads with offline learning.
}
\label{tab:model_comparison}
\end{table}

\section{Problem Statement}

\label{chapter:problem_statement}

\textit{We begin with a formal description of the underlying optimization problem and the intuition behind its main ingredients.}
\subsection{Search Space, $\Theta$}

\begin{figure}
\centering
\includegraphics[width=0.48\textwidth]{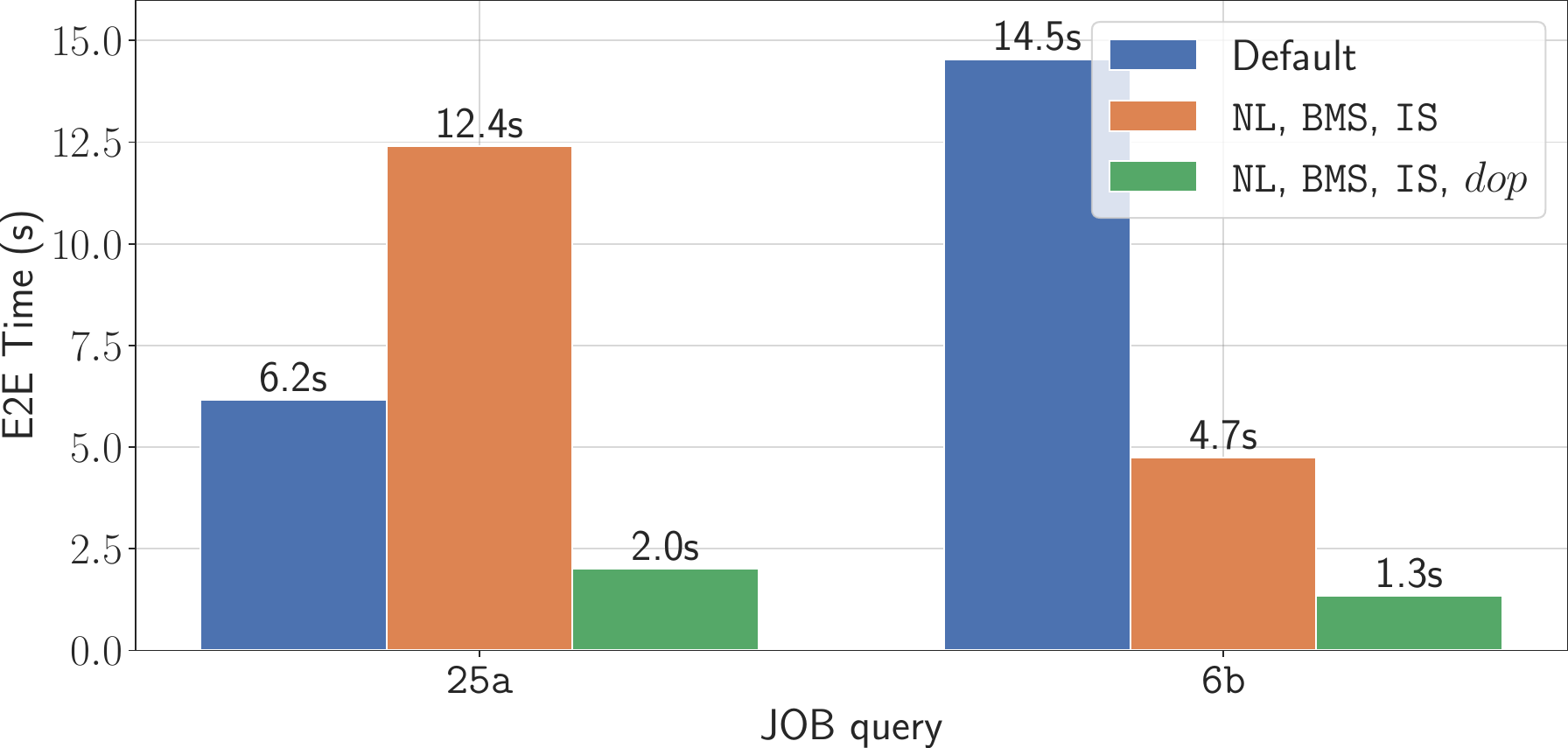}
\caption{Examples where $dop$ control is necessary to \textit{unlock} the potential of operator-related hints (query \texttt{25a}) and to \textit{increase} it (query \texttt{6b}).
}
\label{fig:25a_6b}
\end{figure}

Hints are typically used to guide an optimizer in the search for better plans  \cite{marcus2020bao}, but they can also be used to control operation modes such as the degree of parallelism (query $dop$). For example (Fig. \ref{fig:25a_6b}), for JOB query \texttt{25a}, cardinality underestimation leads to \texttt{INL Join}, causing a significant index lookup overhead. Disabling \texttt{INL Join} leads to a \texttt{Hash Join}, but it is still inefficient due to data broadcast overhead amplified by the default use of 64 threads. Only by reducing $dop$ was the overhead mitigated, \textit{unlocking} the potential of join options. For query \texttt{6b}, the overhead is not significant and controlling $dop$ results in just an \textit{additional} speedup.

\begin{mytakeaway}
    \textbf{Observation (O1):} \textit{Controlling $dop$ may be crucial for unlocking the potential of operation-related hints}. 
\end{mytakeaway}

\begin{table*}[htbp]
\centering
\begin{tabular}{c c c c c c c c c c c c c c c c}
\toprule
\multicolumn{1}{c}{Benchmark} & \multicolumn{3}{c}{Exploration Time (sec)} & \multicolumn{3}{c}{Planning Time (sec)} & \multicolumn{3}{c}{Unique Plan Stuctures} & \multicolumn{3}{c}{Unique Plans} & \multicolumn{3}{c}{Search Space Size} \\
\cmidrule(lr){2-4} \cmidrule(lr){5-7} \cmidrule(lr){8-10} \cmidrule(lr){11-13} \cmidrule(lr){14-16}
 & $\Theta_{dop}$ & $\Theta_{ops}$ & $\Theta$ & $\Theta_{dop}$ & $\Theta_{ops}$ & $\Theta$ & $\Theta_{dop}$ & $\Theta_{ops}$ & $\Theta$ & $\Theta_{dop}$ & $\Theta_{ops}$ & $\Theta$ & $\Theta_{dop}$ & $\Theta_{ops}$ & $\Theta$ \\
\midrule 
TPCH & 1493 & 61156 & 182478 & 0 & 9 & 25 & 35 & 415 & 599 & 35 & 415 & 600 & 66 & 2816 & 8448 \\
JOB & 1445 & 58468 & 172379 & 264 & 7511 & \underline{\textbf{21338}} & 206 & 3837 & 7050 & 215 & 4167 & 7692 & 339 & 14464 & 43392 \\
SQ & 2002 & 75048 & 223740 & 158 & 4804 & \underline{\textbf{13109}} & 95 & 2064 & 4271 & 97 & 2162 & 4524 & 120 & 5120 & 15360 \\
\bottomrule
\end{tabular}
\caption{Even \underline{query planning} becomes impractical due to the rapid expansion of the search space when adding $dop$. However, the low uniqueness of plans gives a \textit{hope} for an efficient exploration by algorithms better than brute force.}
\label{tab:search_space_extension}
\end{table*}

\noindent Motivated by this observation, we extend the operation-related parameter space $\Theta_{ops}$ by parallelism control $\Theta_{dop}$:
$$
\Theta = \underbrace{\Theta_{scans} \times \Theta_{joins}}_{\Theta_{ops}} \times \textcolor{Maroon}{\mathbf{\Theta_{dop}}}.
$$

\begin{figure}
\centering
\includegraphics[width=0.48\textwidth]{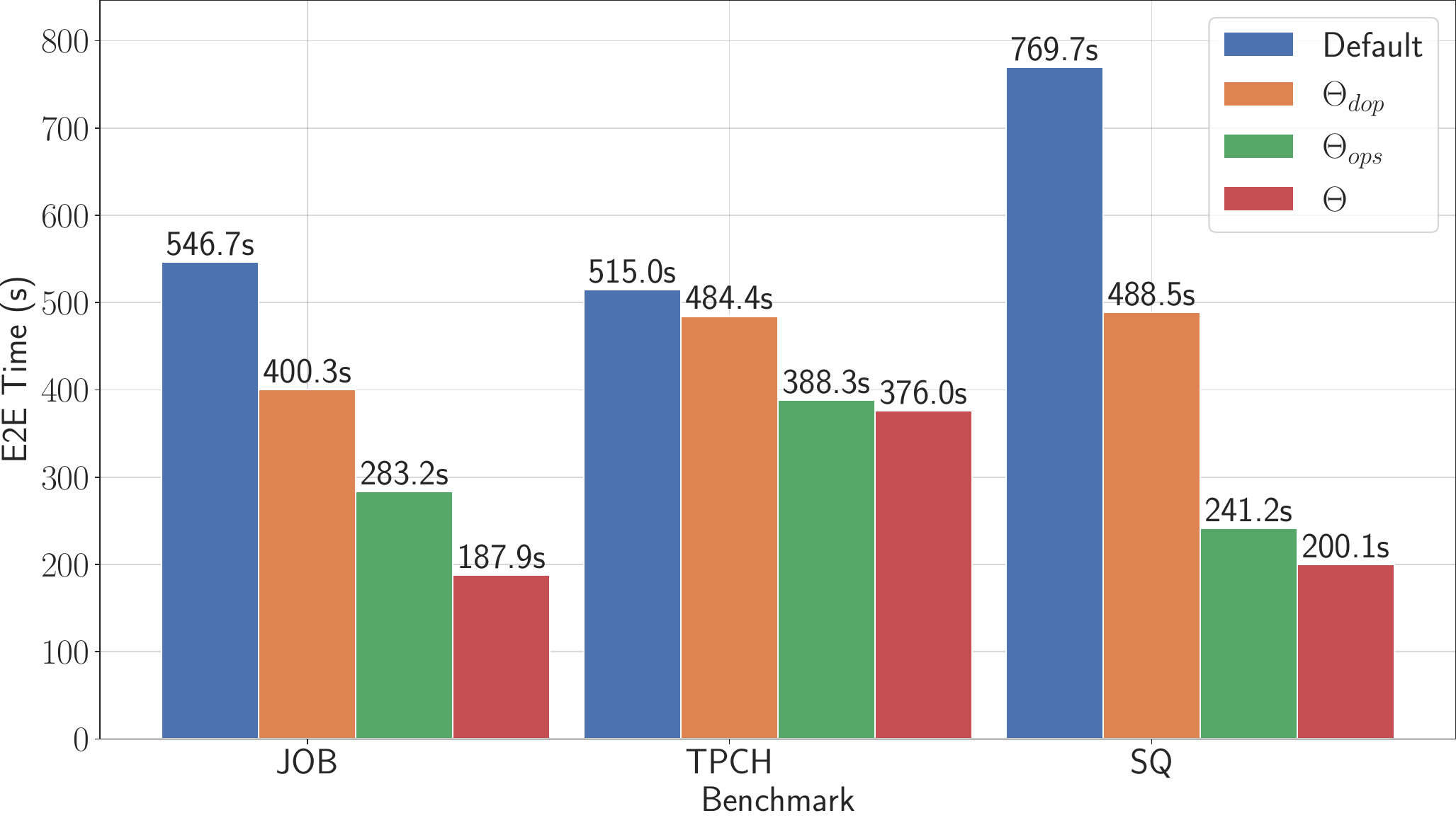}
\caption{Influence of hint types on  benchmark latency.}
\label{fig:search_spaces}
\end{figure}

\noindent This extension significantly enhances the potential of hint-based optimization, increasing the potential boost from 2x to 3x on JOB benchmark (Fig. \ref{fig:search_spaces}) \footnote{and, as we show in experiments, HERO achieves this boost (Tab. \ref{tab:ideal_static_workload})}. Note that improvement depends on a benchmark, with larger gains observed on more challenging workloads as the boost is obtained because of \textit{correcting} planner errors. Obviously with the increase in performance we are also getting increase of the search space, which in turn, requires algorithms for fast training and inference. 

To represent the set of operation-related hints, we fix the order of operators\footnote{in this work, we consider \texttt{Nested Loop Join}, \texttt{Hash Join}, \texttt{Merge Join}, \texttt{Bitmap Scan}, \texttt{Index Only Scan}, \texttt{Index Scan}, \texttt{Seq Scan}} and encode the disabled operators using a \textit{bitmask}. For instance, disabling \texttt{MJ} and \texttt{BMS} is encoded as $\theta_{0\text{b}0011000} = \theta_{24}$.

\subsection{Dependency Under Study}

\noindent A query is processed in two stages: a planner $PL$ creates a plan which is then executed.  The influence of hints on query execution time is limited to changes in the plan which also depends on statistics $Stats$. Thus, the dependency between query latency and hints can be represented as

$$
    t(\textcolor{Maroon}{\boldsymbol{\theta}}) = t^{pl}(q, \textcolor{Maroon}{\boldsymbol{\theta}}, \textcolor{black}{Stats}) + t^{ex}(PL(q_i,\textcolor{Maroon}{\boldsymbol{\theta}}, \textcolor{black}{Stats}), \textcolor{black}{\textbf{\dots}}).
$$

\noindent So, the task of hint-based optimization is to select such parameters $\theta$ that the queries $q_i$, characteristic for a workload $Q$, are executed as quickly as possible:

$$
    t(Q) = \mathbb{E}_{q_i \in \mathcal{P}_{Q}} t_i \rightarrow \min_{\textcolor{Maroon}{\boldsymbol{\theta}} \in \Theta}.
$$

\begin{mytakeaway} \textbf{Observation (O2)}: \textit{Query latency depends on hints \underline{only} via their influence on the plan.}
\end{mytakeaway}

\subsection{Optimization Problem}

\begin{figure*}
\centering
\includegraphics[width=\textwidth]{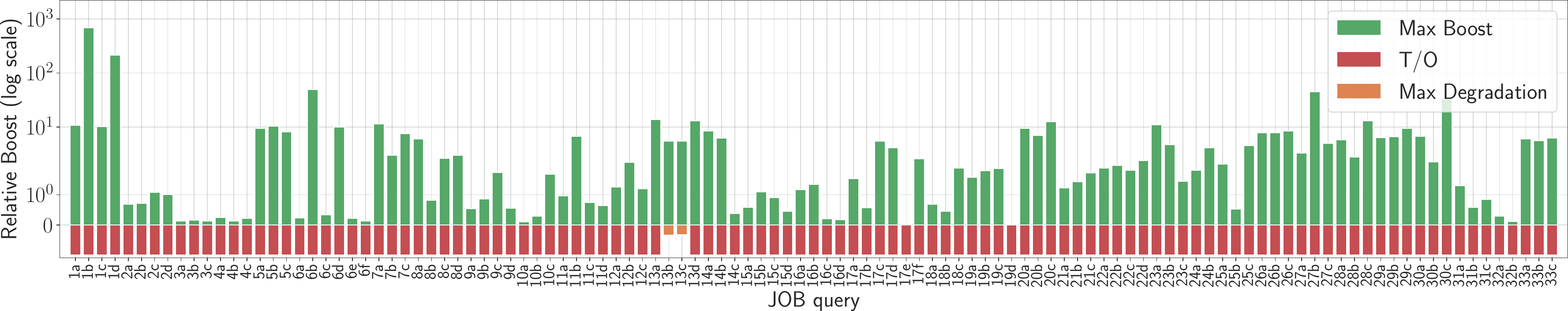}
\caption{Query execution results with the best and worst hints. As one can see, the degradation problem is highly relevant: virtually any query can be significantly slowed down by a wrongly selected hint.}
\label{fig:JOB_extreme_cases}
\end{figure*}

\noindent As one can readily see in Fig. \ref{fig:25a_6b}, there is no single $\theta$ that optimizes all queries. To avoid manual hint selection, we need to build a predictive model $M$ that, given some information $info$ about a query, can predict an optimal (or at least `good') set of hints $\theta^*$. Since we may not know a workload $Q$ in advance, our model must make reliable predictions for \textit{any} query. The results in Figure \ref{fig:JOB_extreme_cases} clearly show the problem of degradation due to improperly selected hints. So, our goal is to build a predictive model $M$ that minimizes the average workload latency while ensuring reliable and useful predictions:

\begin{equation*}
    \begin{aligned}
        &\mathbb{E}_{q_i \sim \mathcal{P}_{Q}} \big[ t_i^M + t_i\vert_{\bm{\textcolor{Maroon}{\theta=M(info_i)}}} \big] \rightarrow \min_{\bm{\textcolor{Maroon}{M}}}, \\
        &\forall q_i \in \text{supp}\big(\mathcal{P}_{Q}\big): t_i^M + t_i\big \vert_{\theta=M(info_i)} \leq t_i \big \vert_{\theta_{def}},
    \end{aligned}
\end{equation*}

\noindent where $t^M$ is the time for executing the model and calculating \textit{info}.

\subsection{Query Representation}
\label{subchapter:query_representation}

\noindent \textbf{Query as a text ($\bm{info = q_i}$).} If the query is represented solely by its syntactic form, ensuring the required level of reliability would require the model to have an understanding of query plans generated with various $\theta$s and their  execution times. Given the inherent complexity of planner and dependency on data statistics, this approach looks  \textit{problematic}. Either the model must have knowledge on planner operation and current statistics, or the training data must be designed in a way to provide this information. The latter option looks highly impractical, considering the exponential number of hint combinations and the dependency of the plan on statistics. Besides, this approach is complicated by the problem of data privacy, because raw query text may disclose vulnerable information contained in query predicates.

\noindent \textbf{Query as a plan ($\bm{info = plan_{def}}$).} Intuitively, traditional query optimizers incorporate decades of expertise, which may be used by the model. However, representing a query solely by its default plan is problematic: as shown in Table \ref{tab:plan_representation}, predictions based only on the default plan are not reliable.

\begin{table}
\centering
\resizebox{\columnwidth}{!}{%
\begin{tabular}{|c|c|c|c|}
\hline
\textbf{JOB Query} & \textbf{Def. Time (sec)} & \textbf{Custom Time (sec)} & \textbf{Boost} \\
\hline
\texttt{6b} & 14.53 & 8.95 & x1.62 \\
\hline
\texttt{6d} & 14.36 & \textcolor{Maroon}{\textbf{T/O}} & \textcolor{Maroon}{\textbf{NaN}} \\
\hline
\end{tabular}
}
\caption{Queries with the same default plan exhibit significantly different behaviors under the same hint $\theta_{\textbf{ops}} = 126$.}
\label{tab:plan_representation}
\end{table}

\begin{figure}
\centering
\includegraphics[width=0.48\textwidth]{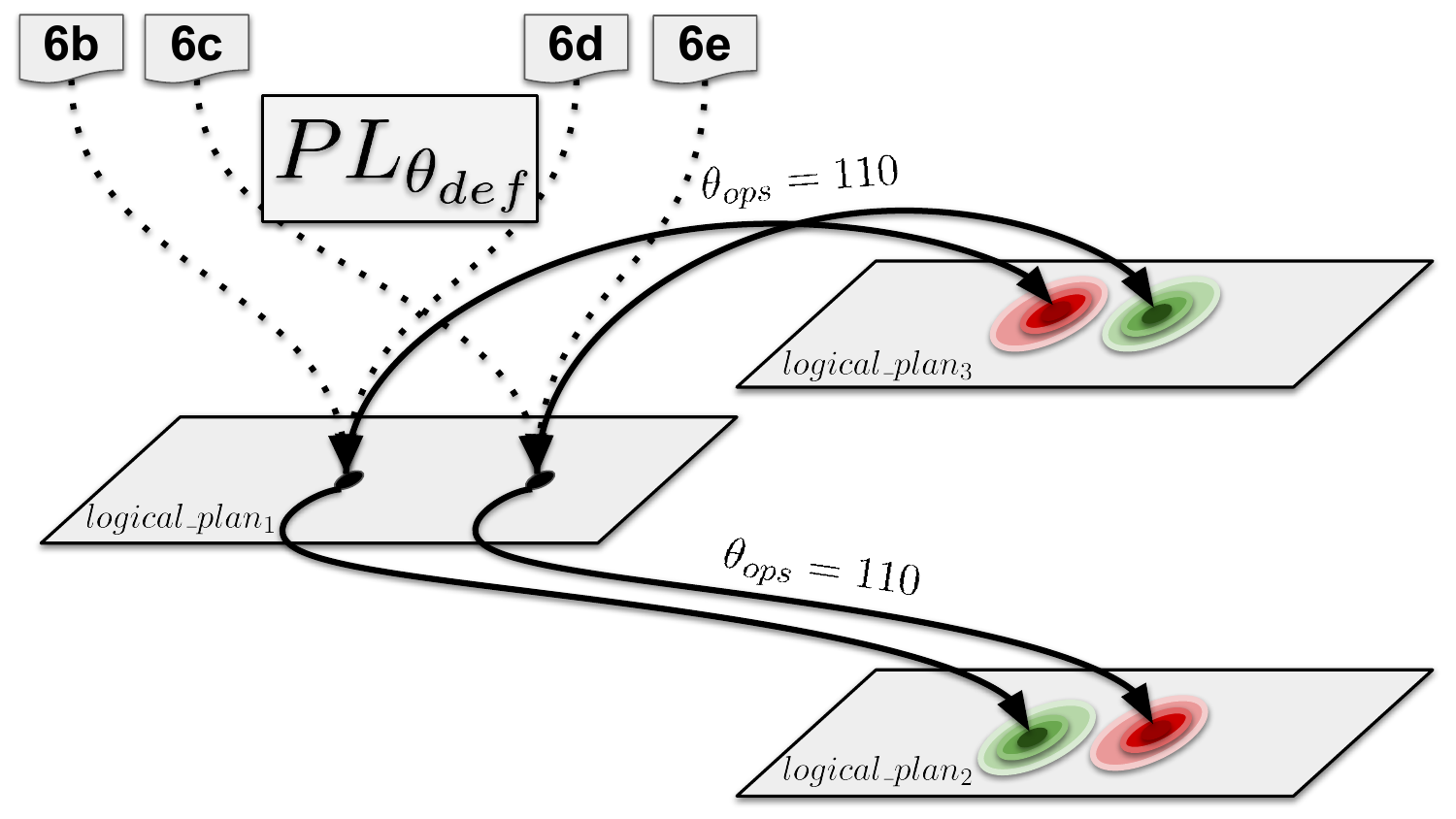}
\caption{Due to  \textit{collisions} it is difficult to reason about query acceleration in the plan space without having information about the underlying queries.}
\label{fig:6bcde}
\end{figure}

\noindent \textbf{Query as a \textit{set} of plans ($\bm{info = \{plan_j\}}$).}
In Bao \cite{marcus2020bao}, the need to emulate planning dependencies was \textit{completely eliminated}. Instead, the model just approximates the dependency \( plan \rightarrow t^{ex} \) without any knowledge about hints or the planning process itself. However, this introduces two challenges. The first challenge is the computational cost: generating $info$ as \(\{ plan_j \}_{j=1}^{|\Theta|} \) requires building an exponential number of plans. With a sufficiently big number of hint options this makes model inference \textit{impractical}. The second challenge is the collection of a consistent and representative dataset which is resource-consuming a non-trivial if one wants to ensure generalizability. Additionally, this method of query representation has the following problem (see Fig. \ref{fig:6bcde}). Pairs of JOB queries such as, e.g., \texttt{6b} and \texttt{6d}, \texttt{6c} and \texttt{6e}, initially share the same default plans. However, after applying the hint, the plan structure changes, causing \texttt{6b} and \texttt{6c} to share the same structure, as do \texttt{6d} and \texttt{6e}. Despite this, only \texttt{6b} and \texttt{6d} are accelerated, while \texttt{6c} and \texttt{6e} experience timeouts. As a result, we have two difficulties at once. First, we cannot unambiguously predict the behavior of a query from its default plan. Second, we cannot reliably estimate the performance of a custom plan without information about which query it comes from. The latter, as we will find out next, is one of the reasons for the complex optimization landscape, which in turn leads to unreliable neural network predictions. The cause of both of these problems will be referred to as the \textbf{plan collision challenge}, thereby weakening the condition of complete plan matching to similarity. We discuss our way of solving this challenge in more detail in \S\ref{subchapter:query_clusterisation}.

\begin{mytakeaway}
\textbf{Observation (O3):} \textit{The choice of \textbf{$info$} is crucial and it can either simplify or complicate the hint prediction task for a model.}
\end{mytakeaway}

\section{HERO's Design}
\label{chapter:hero_design}

\noindent \textit{In this section, we describe the  ideas in the architecture of HERO that overcome the key problems of NN-based solutions (argued further in detail in \S\ref{chapter:justification}).} 
The design of HERO is driven by the following key features/requirements:
\begin{enumerate}[label=\textbf{(R\arabic*)}, leftmargin=*, itemsep=0.3em]
\item degradation risk management
\item near optimal latency acceleration
\item fast model training
\item flexibility to balance between query acceleration and fast training
\end{enumerate}

\label{subchapter:transitions_graph_storage}

\begin{figure*}
\centering
\includegraphics[height=0.25\textheight]{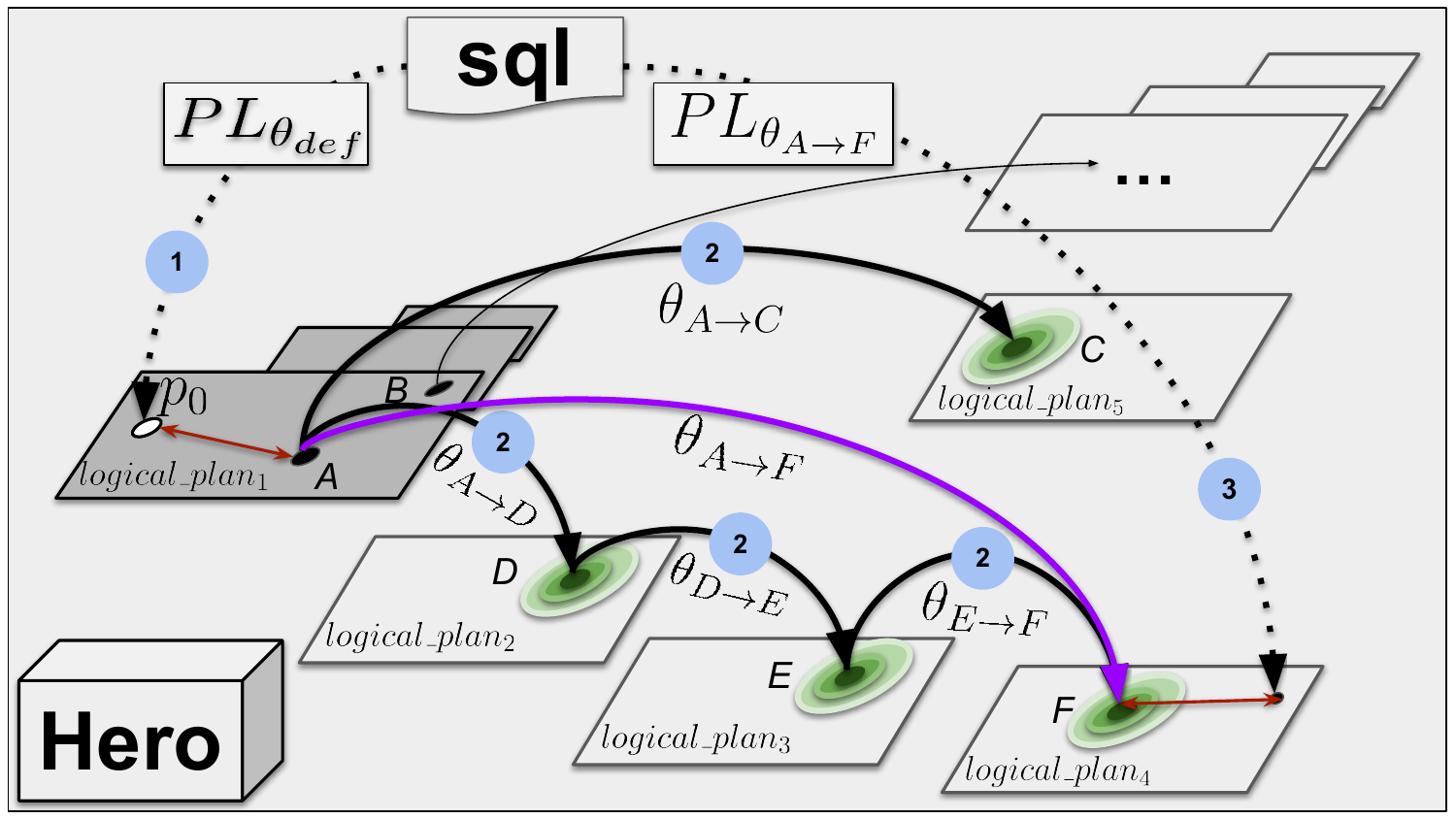}
\caption{By replacing NN with an ensemble of context-aware models, HERO accelerates inference by reusing plan information of similar queries encountered in the past to avoid multiple planning. 
}
\label{fig:graph_storage}
\end{figure*}

\subsection{Query Clusterisation: Contextual Predictions}
\label{subchapter:query_clusterisation}

Example in Fig. \ref{fig:6bcde} shows that in order to guarantee requirement \textbf{(R1)} one has to consider the problem of plan collisions. Interestingly, in order to cope with that problem, we propose to \textit{turn plan collisions into advantage} by clustering queries with the same default plan and by approximating the following \textit{conditional} dependency: 
$$
    \langle plan_0, plan_j \rangle \rightarrow t^{ex}(plan_j)|_{plan_{def}=plan_0}.
$$ 
Firstly, this allows us to take into account the actual statistics, since their impact on the planner behavior will be reflected in the plan. Secondly, this allows for sharing information between queries that look similar from the planner's perspective.

\subsection{Removing NN: Ensemble and Increased Reliability}
\label{subchapter:removing_nn}

To approximate this dependency, we developed an approach conceptually reminiscent to FastGres \cite{woltmann2023fastgres}, resulting in an \textit{ensemble of context-aware models}. Each model in the ensemble handles predictions for queries with the same default plan $p_0$. It contains information on \textbf{(a)} custom plans, \textbf{(b)} hints used to generate them, and \textbf{(c)} observed performance gains for queries with the same $p_0$. This design, combined with a specialized distance metric, allows us to \textit{control the reliability of predictions}. The metric is defined as follows: plans are considered to be infinitely distant if their logical structures differ; otherwise, the distance equals to the relative difference between the vectors of statistics of these plans. This metric is used at two stages. First, this metric is used to determine the most relevant context-aware model from an ensemble by measuring the distance between the default plan of the current query and the plans associated with models. Second, the metric is used to assess custom plans generated by using hints. To ensure reliability \textbf{(R1)}, the model avoids prediction if the distance between the plans exceeds a certain threshold. This means that the model’s predictions rely on reusing information only from a query that \textbf{(a)} had the same logical and default plans, and \textbf{(b)} occurred under similar conditions (statistics) from the optimizer perspective. 

\noindent \textbf{Note on Debugging.} By making the prediction process transparent, we achieve an interpretable model that is easy to debug, which is crucial for integration into production systems. The design allows us to identify exactly which element of the ensemble was used for prediction and understand why it evaluated the plan as it did. As a result, HERO can be deployed even in challenging scenarios. In cases of performance degradation, one can easily \textit{disable} any model that negatively impacts the system, while still utilizing the rest of the ensemble. This option is particularly useful when dealing with sudden changes in workload or environment, as well as with inconsistent or incomplete training data.

\subsection{Hints Prioritization: Fast Inference.} 
\label{subchapter:hints_prioritization}

The main issue with long inference times in existing hint-based optimizers is the need to generate a multitude of query plans for different hints. HERO solves this problem by using a model ensemble that leverages historical data to select promising hints \textit{without requiring multiple planning iterations}. We organize the ensemble as a graph storage, where vertices represent observed plans and edges indicate changes due to hint application and the corresponding performance boosts (see Fig. \ref{fig:graph_storage}). In this representation, data for the context-aware model associated with a plan $\boldsymbol{p}$ corresponds to a subgraph induced by traversing all edges from vertex $\boldsymbol{p}$. So, a key feature of our data is that we store directly obtained prior plans and use them to avoid unnecessary planning calls. The inference procedure is as follows:

\begin{itemize}
    \item First, we generate the default plan ($\boldsymbol{p_0}$) for the given query and identify the closest context-aware model by finding plans similar to it in the graph storage.
    \item Second, we traverse relevant edges to find the most promising hints for planning, notably \textit{without the need to call the planner at each step}, which greatly speeds up the procedure.
    \item Third, after identifying the best hints, we plan the query with them.
    \item And, finally, we assess the resulting plans ($\boldsymbol{p_{\theta_{A \rightarrow F}}}$) to ensure the reliability of the selected hints.
\end{itemize}

\noindent \textbf{Note on super-fast inference.} Our model also supports \textit{super-fast} inference, in which we use top-level clustering based on the query \textit{template}, which can be calculated even \textit{without} planning. Based on it, we can search for the most promising hints among \textit{all} default plans of queries with the same \textit{template} and plan all of them in a first batch along with the default hintset. Until our storage has grown, the number of such promising hints is small enough to put them all into a \textit{single} batch with \textit{parallel} planning.

\subsection{Semantics-Informed Search: Rapid Convergence to Optimum in Extended Search Space}

To satisfy the performance requirement (\textbf{R2}), we propose utilizing both operation-based hints and $dop$ control. While this approach significantly improves query latencies, it also rapidly increases the search space. Our experiments indicate that a simple greedy algorithm (see \S \ref{subchapter:inference} for a description) struggles with search in this space, often selecting suboptimal hint combinations (\S\ref{subchapter:static_workload}). To combat this, we propose to bring knowledge about the semantics of operations inside the algorithm. Firstly, we propose to adjust $dop$ simultaneously with settings for join-hints. Secondly, we use planner-specific combination of hints. For example, to turn off \texttt{Index Nested Loop Join} in openGauss we need to disable \texttt{BitMap Scan} and \texttt{Index Scan} as well as \texttt{Nested Loop Join}.

\subsection{Parameterized Local Search Procedure: Balancing Between Performance and Exploration}

To meet requirements \textbf{(R3)} and \textbf{(R4)}, we propose a parameterization of the semantics-informed search algorithm that adjusts the changes made at each iteration. This approach seamlessly integrates into a local search framework, where the parameters define the size and structure of local neighborhoods, guiding the exploration process and enabling the algorithm to transition to more promising states when improvements are detected. The design of our parameterization rests on several intuitive assumptions about the search space:

\begin{enumerate}[label=\textbf{(A\arabic*)}, leftmargin=*, itemsep=0.3em]
    \item Join strategies generally have a larger impact than table scan operators.
    \item Simultaneously selecting the join type and degree of parallelism ($dop$) is highly efficient.
    \item Only minor adjustments to the default plan are often sufficient, meaning a few iterations may yield significant improvements.
    \item The most common openGauss mistake is related to applying  \texttt{Index Nested Loop Join} to large relations.
\end{enumerate}  

\noindent For each of them, we introduced a different parameter that allows us to appropriately influence the shape of the neighborhood during local search. The effectiveness of this parameterization was experimentally confirmed in \S\ref{subchapter:parameterization}.

\subsection{Ranking over Regression: Accelerating Exploration}

\noindent Training HERO essentially involves collecting information for graph storage. Due to easy debugging and the ability to disable context-aware models, as discussed in \S \ref{subchapter:removing_nn}, HERO has lower data quality requirements compared to other solutions. This allows us to speed up training, e.g., by prioritizing the preservation of ranking between plans (we discuss its importance in \textbf{(O4)}) rather than focusing solely on the accuracy of predicted execution times. To achieve this, we use a parallelized local search where all neighboring plans are explored (executed) at each iteration  \textit{offline}\footnote{offline learning allows us to avoid degradation issues in user space and achieve a higher latency boost compared to online learning \S \ref{subchapter:static_workload}} and \textit{simultaneously}. Parallelization significantly reduces exploration time and in our experiments in a real environment we observed that the plan selected from parallel execution is often either the fastest one or very close to the optimal plan obtained from sequential execution. To further ensure the reliability of the ranking, periodic \textit{sequential pairwise comparisons} between candidates can be performed during the exploration phase for validation.

\section{Design of Existing NN-Based Solutions}
\label{chapter:pipeline}

\begin{figure*}
\centering
\includegraphics[height=0.3\textheight]{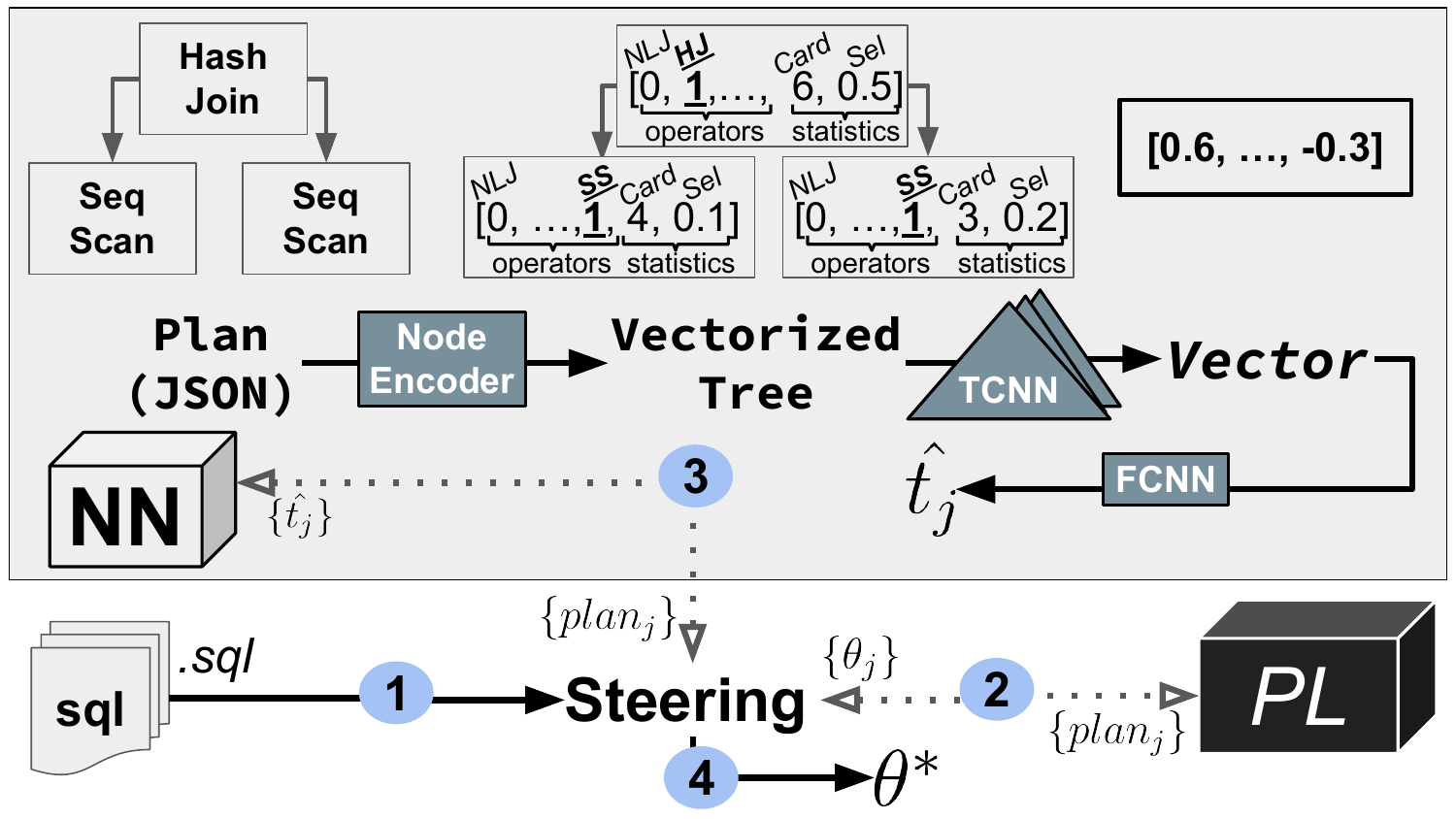}
\caption{Typical inference schema in existing NN-based solutions for hint-based query optimization.}
\label{fig:inference}
\end{figure*}

\textit{In the existing approaches to hint-based query optimization, regression ML models are typically used to predict plan execution time. They are combined with a  \texttt{Steering} module responsible for exploring hint combinations and generating corresponding plans (Fig. \ref{fig:inference}). In this section, we analyze the main steps of this pipeline and in the subsequent section we analyze its limitations.}

\subsection{Model}

\textbf{Encoding.} One of the main challenges in using NNs for predicting plan properties lies in the fact that  plans are structured objects of varying lengths, whereas conventional machine learning algorithms operate on fixed-size vectors. Moreover, each node within an execution plan contains significant information (e.g., statistics) that needs to be taken into account. A common approach is to use one-hot encoding for operations, along with additional channels representing statistics, such as estimated costs, cardinalities, and computed selectivities. However, if hints in a DBMS are implemented by \textit{heavily overestimating costs} of alternative plans (this is the case for openGauss \cite{opengauss} and PostgreSQL \cite{postgresql}, for instance) then it makes no sense to integrate costs into plan representation. Different options for node encoding are discussed in detail in \cite{zhao2023comparative}.

\noindent \textbf{Tree Vectorization.} To vectorize the tree structure of plans, in \cite{mou2016convolutional} the authors proposed a special convolutional layer which is a variant of a graph convolutional network. Since query execution plans generally have at most two child nodes, this layer can be simplified to a 1-D convolution, similar to those used in image processing. This simplification enables the use of highly efficient techniques from image analysis. Combining these layers with non-linear transformations gives convolutional blocks. We refer to a sequence of such blocks as a Tree Convolutional Neural Network (TCNN) \cite{mou2016convolutional}. After passing the tree through it, the hierarchical structure of subtrees is captured, and the tree can be easily vectorized using any aggregation function. The resulting vector can then be fed into a classical Fully Connected Neural Network (FCNN). In addition to having an efficient implementation, this idea also introduces an inductive bias into the model, because we `assist' the model by embedding our prior knowledge about the data structure directly into the architecture. We note that this concept is actively developed in the field of Geometric Deep Learning \cite{bronstein2017geometric}. By exploiting the inherent geometry of the data, the model is able to generalize more efficiently and learn more meaningful representations, as we demonstrate in \S\ref{chapter:justification}.

\subsection{Inference}
\label{subchapter:inference}
The inference process in the existing NN-based solutions is implemented by a \texttt{Steering} module and looks as follows:
\begin{itemize}
    \item First, it produces execution plans for different hint combinations ${\theta_j}$.
    \item Second, \texttt{Steering} feeds them into NN to predict execution time $\hat{t_j}$.
    \item Third, based on the obtained predictions, it selects the most promising hint set.
\end{itemize}
\noindent In general, the overall process can be iterated, refining hint suggestions and generating new plans for evaluation. We provide an illustration in Figure \ref{fig:inference}. 

\noindent The key point in the inference process is the need to generate and evaluate plans for different hint combinations which leads to the challenge \textbf{(C1)} of trade-off between two competing goals: finding hints $\theta^*$ that lead to an optimal plan while minimizing the number of combinations explored, i.e., restricting exploration to a subset $\{\theta_j\} \subseteq \Theta$. In Bao \cite{marcus2020bao}, an expensive \textit{exhaustive} search algorithm was used. Later, in AutoSteer \cite{anneser2023autosteer}, the authors proposed to leverage the structure of  hint space, noting that hints from $\Theta_{ops}$ impose a partial order on  operators. This allowed them to use a \textit{greedy algorithm} to search for the best hint combination. However, the greedy approach has drawbacks: it is not guaranteed to find the optimum and its sequential nature makes it inefficient for parallelization (as the size of the hint space grows, the overhead becomes prohibitive).

\begin{mytakeaway}
\textbf{Observation (O4):} \textit{For inference, plan ranking is more important than  predicted execution times.}
\end{mytakeaway}

\subsection{Training and Data Labeling}

\noindent Default plans produced by a DB optimizer is not a sufficient source of information for training a good model, as it fails to provide information about the behavior of non-default plans.\footnote{however, as shown by the results of the 0th epoch in Fig. \ref{fig:all_dops_online_job}, even this limited information is sufficient for NN-based solutions with the pruned inference algorithm to speed up workload} Thus, an \textit{exploration} stage is required to analyze query latency under different hint combinations. Exploration can be made \textit{offline} to avoid runtime degradations or \textit{online} to account for data/performance drift. Exploration is costly: we have to wait for the query to execute with each of the hint combinations of interest. Moreover, some hints can slow down queries by orders of magnitude, making labeling \textit{prohibitively} expensive. Efficient exploration is the second principle challenge \textbf{(C2)} in learned hint-based optimization. To solve this problem, we can narrow down the exploration area similarly to the greedy algorithm used at the inference stage. However, as we found out in practice, the greedy algorithm misses good solutions and does not allow for unlocking the full potential of optimization with hints.

\section{On the Reliability of NNs}
\label{chapter:justification}
\noindent \textit{In this section, we provide results of an experimental study on  limitations of NN models for hint-based optimization.}

\subsection{Experimental Setup}

\textbf{Data Split.} Here and later in the paper we use queries from the IMDb dataset (JOB and SQ). Initially, we grouped all plans for JOB queries under all possible hint combinations, splitting them 4:1 into train and validation subsets randomly. Additionally, we split SQ queries into two groups: those with default logical plans matching JOB queries (Test) and those without (out of distribution, OOD). This allowed us to reflect the similarity of the queries with those from JOB. All learning curves and metrics mentioned in this section are averaged over five runs.

\begin{figure}
\centering
\includegraphics[width=\columnwidth]{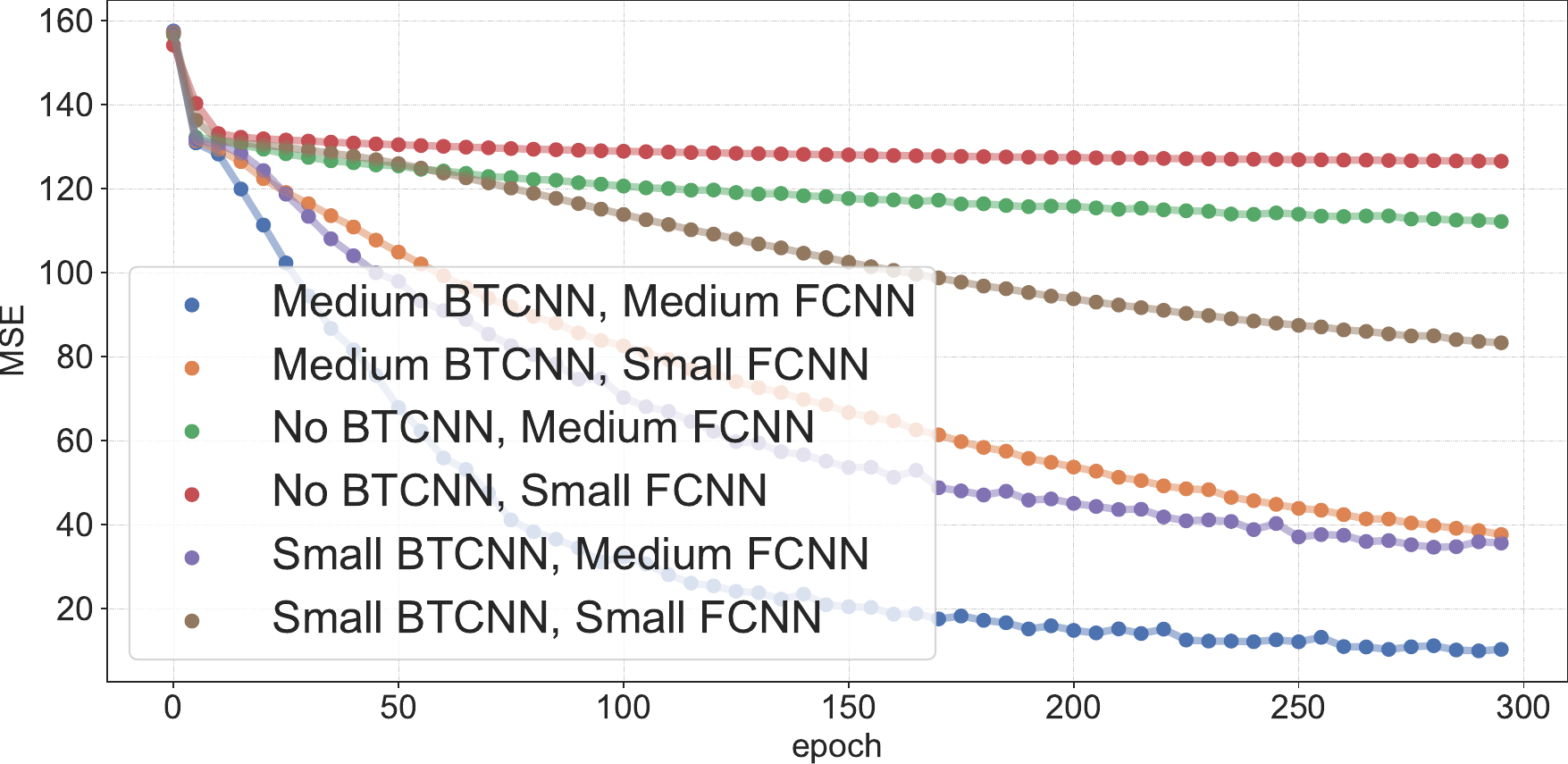}
\caption{Learning dynamic on the training data under different TCNN and FCNN block ratios demonstrates that TCNN efficiently learns a good representation of plans, which simplifies the optimization problem by accelerating error reduction.
}
\label{fig:medium_models}
\end{figure}

\noindent \textbf{Baseline.} 
In our experiments, we used NN architectures similar to those in Bao \cite{marcus2020bao} and AutoSteer \cite{anneser2023autosteer}. As shown in Fig. \ref{fig:medium_models}, experiments with varying ratios of TCNN and FCNN blocks demonstrate that adding TCNN greatly improves plan vectorization and simplifies prediction tasks. Additionally, we observed that tree normalization layers play a critical role in building an efficient architecture. Considering the semantics of our data, we implemented and applied a custom variant of \texttt{InstanceNormalization}. After all these experiments, we settled on \textit{larger} TCNN and FCNN block sizes, which reduced the error on the training set to zero.

\noindent \textbf{Visualization.} To understand model behavior, we visualized points with high prediction errors (Fig. \ref{fig:extreme_errors}). Plans were processed through a TCNN block to generate fixed-dimensional vector representations. These representations were projected onto a 2D plane, with noise added to the embeddings for data augmentation before constructing the approximation surface.

\subsection{Insights}

\begin{figure}
\includegraphics[width=\columnwidth]{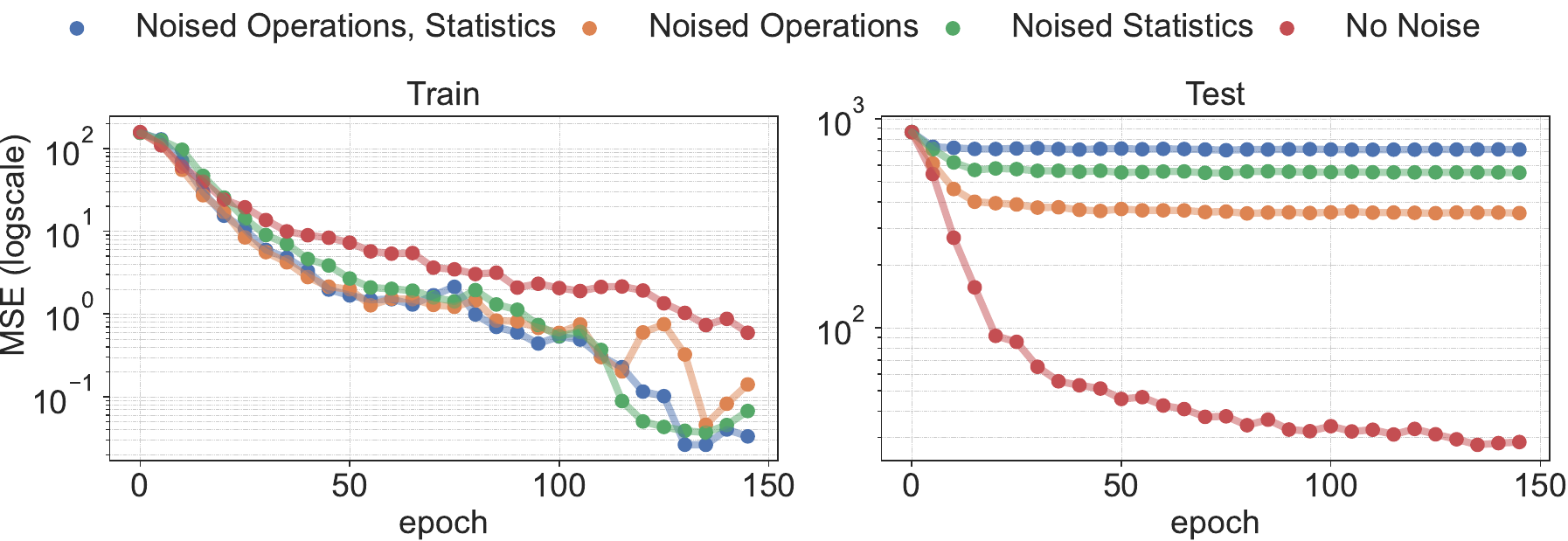}
\caption{Using noise instead of plan node information simplifies the optimization task, increasing the speed of data memorization by an order of magnitude. However, it also reduces the generalization capability.}
\label{fig:noised_data}
\end{figure}

\textbf{Noised Data.} To identify which features are most important and what NN relies  on in making predictions, we set the following experiment. We considered three types of noised plans: \textbf{(a)} noise instead of information about operators in plan nodes, \textbf{(b)} noise instead of node statistics, and \textbf{(c)} fully noised nodes. The results shown in Figure \ref{fig:noised_data}, 
confirm our intuition that statistics encapsulate optimizer's `knowledge' into a compact form and are an important feature. For example, by relying on cardinality, NN in fact, becomes a learned cost model. However, due to the different asymptotic behaviors of operators, NN struggles with situations when noise replaces information about them. 

\begin{myinsight}
    \textbf{Insight (I1):} \textit{Statistics implicitly represent the logic of DB optimizer and are an important feature, positioning NN as a \underline{cost model} with time as the unit.}
\end{myinsight}

\begin{myinsight}
    \noindent \textbf{Insight (I2):} \textit{Plan collisions significantly complicate the optimization landscape so that \underline{even a uniform noise}, playing the role of an artificial labeling, helps to simplify optimization.} 
\end{myinsight}

\begin{figure*}
\includegraphics[width=\textwidth]{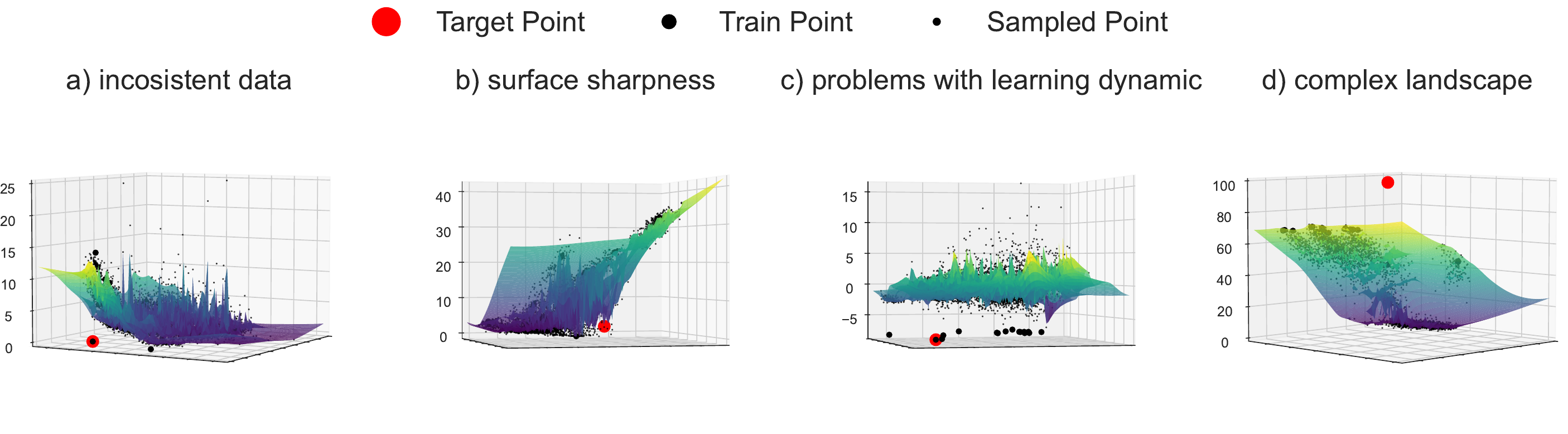}
\caption{Four key challenges in training a reliable regressor. Overestimations can occur due to: \textbf{a)} data inconsistency caused by optimistic execution time labels (e.g., when long-running plans are interrupted), and \textbf{c)} difficulties in learning very small values due to the model architecture (note that y axis is in logscale). Underestimations arise from: \textbf{b)} the sharpness of the surface, which is difficult to approximate even when having sufficient data, and \textbf{d)} the complex landscape of the function, where gathering information on all peaks is challenging due to their abundance and steepness.}
\label{fig:extreme_errors}
\end{figure*}

\medskip

\noindent \textbf{Train Data}. Using the timeout value as a label for degraded queries resulted in a \textit{optimistic} labels of execution time. As a consequence, some of NN predictions started to look like an overestimation error (Fig. \ref{fig:extreme_errors}a). In a detailed analysis we found that the model was correct: it identified a trend by `looking' at neighboring plans with significantly higher execution times and produced predictions \textit{close} to the \textit{real} execution time. But we note that in practice there may not be enough information about informative neighboring plans.

\begin{myinsight}
    \textbf{Insight (I3):} \textit{Due to the need to represent knowledge of all plans in the weights of a \underline{single NN}, any inconsistency (inevitable due to challenge \textbf{(C2)}) may lead to error.} 
\end{myinsight}

\medskip

\noindent \textbf{Learning Dynamics.} We also observed a consistent overestimation of plans with low values, even when there were no nearby high-value plans. This likely stems from the weight learning dynamics, particularly due to the use of the $softplus$ function. In regions where predictions are around 200ms, the gradient during backpropagation decreases by roughly a factor of ten.  As a result, for later training stages, the small learning rate makes it difficult to adjust weights in these regions.

\medskip

\begin{figure}
\includegraphics[width=\columnwidth]{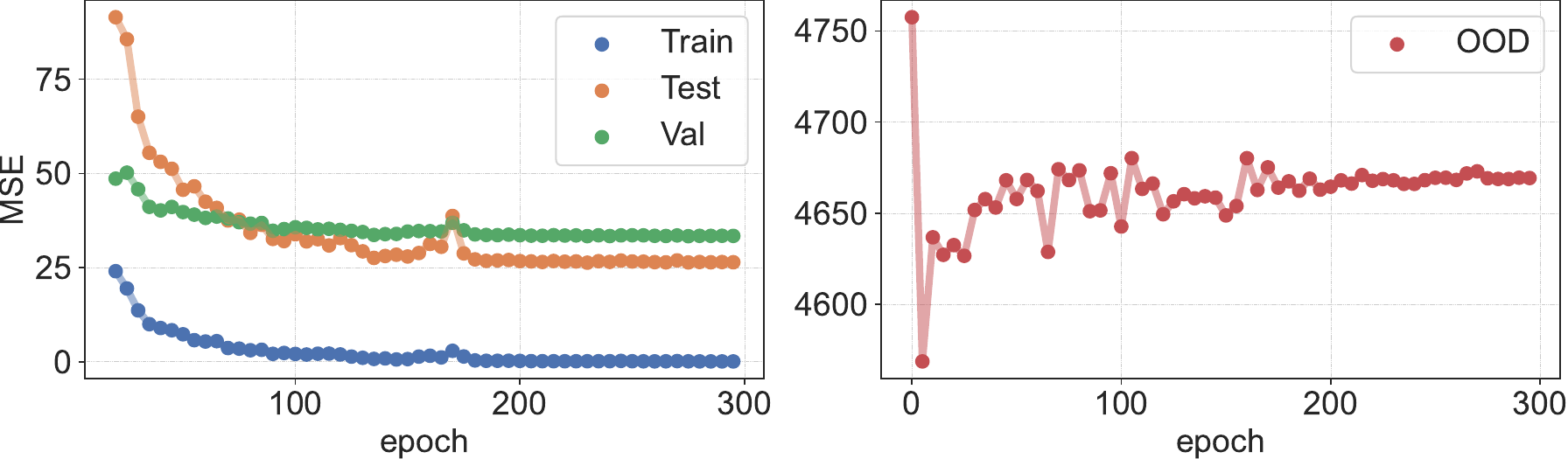}
\caption{The model successfully memorizes the training data and is able to generalize to some extent on validation and test sets, but with a significantly higher error. Generalization on the test set is better than on the validation one, likely due to the structure of plans. At the same time, there is practically no generalization on OOD.}
\label{fig:generalisation}
\end{figure}

\begin{figure}
\centering
\includegraphics[width=\columnwidth]{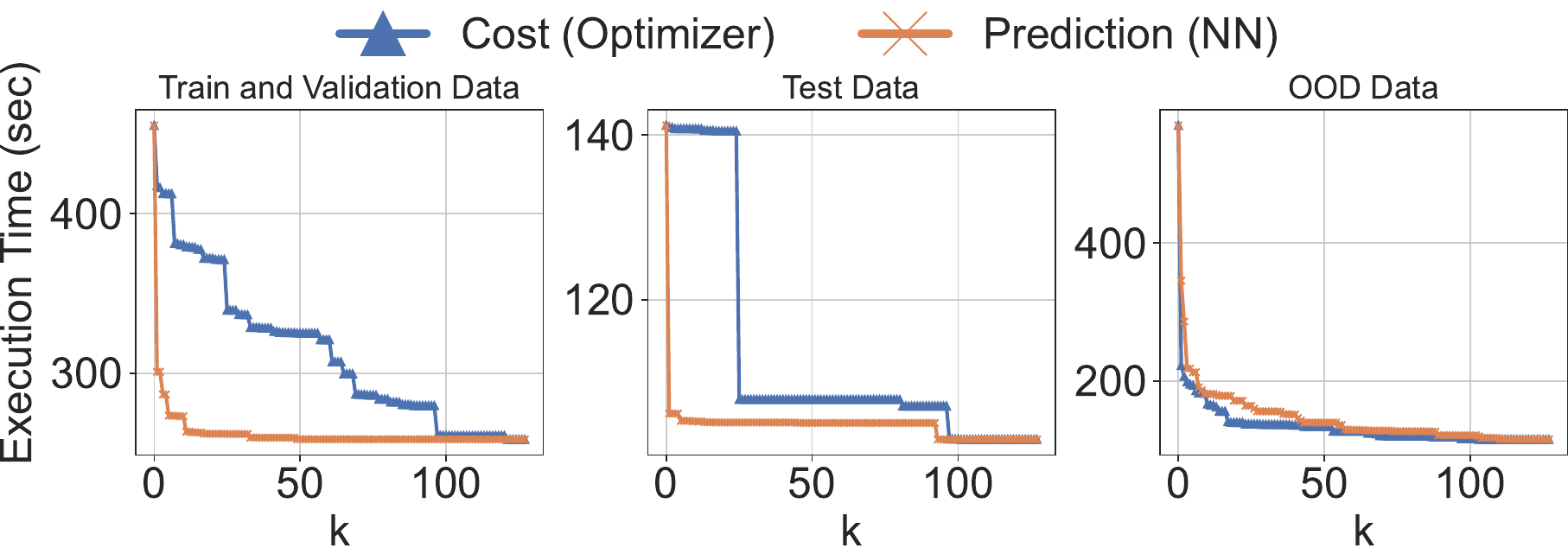}
\caption{Performance of top-k plans based on NN predictions and optimizer costs demonstrates that NN can efficiently solve the plan ranking problem.}
\label{fig:ranking}
\end{figure}

\noindent \textbf{Validation Data.} 
The real challenges of generalization become apparent when analyzing errors on plans not included in the training set (confirming our intuition about the complex nature of the approximated surface). We observed regions with steep slopes, where a lack of information led to overestimations, and flatter areas with isolated peaks, resulting in underestimations due to sparse data (see (b) and (d) in Fig. \ref{fig:extreme_errors}). Despite validation errors were noticeably higher than those on the training data (Fig. \ref{fig:generalisation}), the model was still able to capture key patterns in predicting execution times. This led us to the following question: are NN predictions at least more reliable than optimizer cost estimates? 
To answer this, we first measured two types of correlations between neural network prediction / optimizer cost and query runtime. The results are presented in Table \ref{tab:correlation_metrics}. Observing that neural network predictions are often more correlated with the actual query execution times, we decided to compare the performance of the best plan selected from top-k candidates ranked by either optimizer costs or NN-predicted values (Fig. \ref{fig:ranking}).
As a result, we see that NN can indeed be effectively used for ranking tasks on validation and test data, although no such advantage is observed on the OOD workload.

\begin{table}
\centering
\resizebox{\columnwidth}{!}{
\begin{tabular}{c c c c c}
\toprule
& \multicolumn{2}{c}{\textbf{Kendall}} & \multicolumn{2}{c}{\textbf{Spearman}} \\
\cmidrule(lr){2-3} \cmidrule(lr){4-5} 
\textbf{Data} & \textbf{Cost} & \textbf{Predictions} & \textbf{Cost} & \textbf{Predictions} \\
\midrule
Train  & 0.061$\pm$0.000 & 0.856$\pm$0.009 & 0.076$\pm$0.000 & 0.954$\pm$0.006 \\
Val   & 0.153$\pm$0.024 & 0.723$\pm$0.019 & 0.213$\pm$0.034 & 0.865$\pm$0.016 \\
Test    & 0.199$\pm$0.005 & 0.895$\pm$0.007 & 0.277$\pm$0.008 & 0.974$\pm$0.003 \\
OOD   & 0.161$\pm$0.000 & 0.184$\pm$0.016 & 0.242$\pm$0.000 & 0.271$\pm$0.024 \\
\bottomrule
\end{tabular}
}
\caption{Correlations between NN prediction / optimizer cost and query execution time show that a well-trained neural network can rank plans much better than the optimizer.}
\label{tab:correlation_metrics}
\end{table}

\begin{myinsight}
\textbf{Insight (I4)}: \textit{The \underline{complex geometry} of the studied surface challenges reliable generalization, yet NN is able to outperform DB optimizer in ranking accuracy, provided there is no distribution shift.}
\end{myinsight}
 
\medskip

\noindent \textbf{Test Data.} Analyzing the behavior on the test dataset (in which plans come from queries not present in the training data), we observed an interesting phenomenon: the errors were significantly lower than those on the validation dataset (Fig. \ref{fig:generalisation}). To understand it, we need to recall that we split plans from SQ into OOD and Test based on their logical structure and consider how TCNN processes tree structures. We computed the ratio of subtree structures from the Validation and Test sets that matched substructures present in the Train. 
We can observe a general trend -- the more complex substructures of plans we consider, the greater the advantage of test data over validation data. For instance, for subtrees of height four, we found that the match rate was approximately $91\%$ for Test, which was twice as high as for Validation.\footnote{we did not consider the other values because the depth of our TCNN is four and it does not learn structural patterns of larger size.} As a result, we conclude that this is one of the reasons for the better generalization.

\begin{myinsight}
\textbf{Insight (I5)}: \textit{The degree of matching between plan substructures and those present in the training set is a signal of the good generalization.}
\end{myinsight}

\begin{figure}
\centering
\includegraphics[width=\columnwidth]{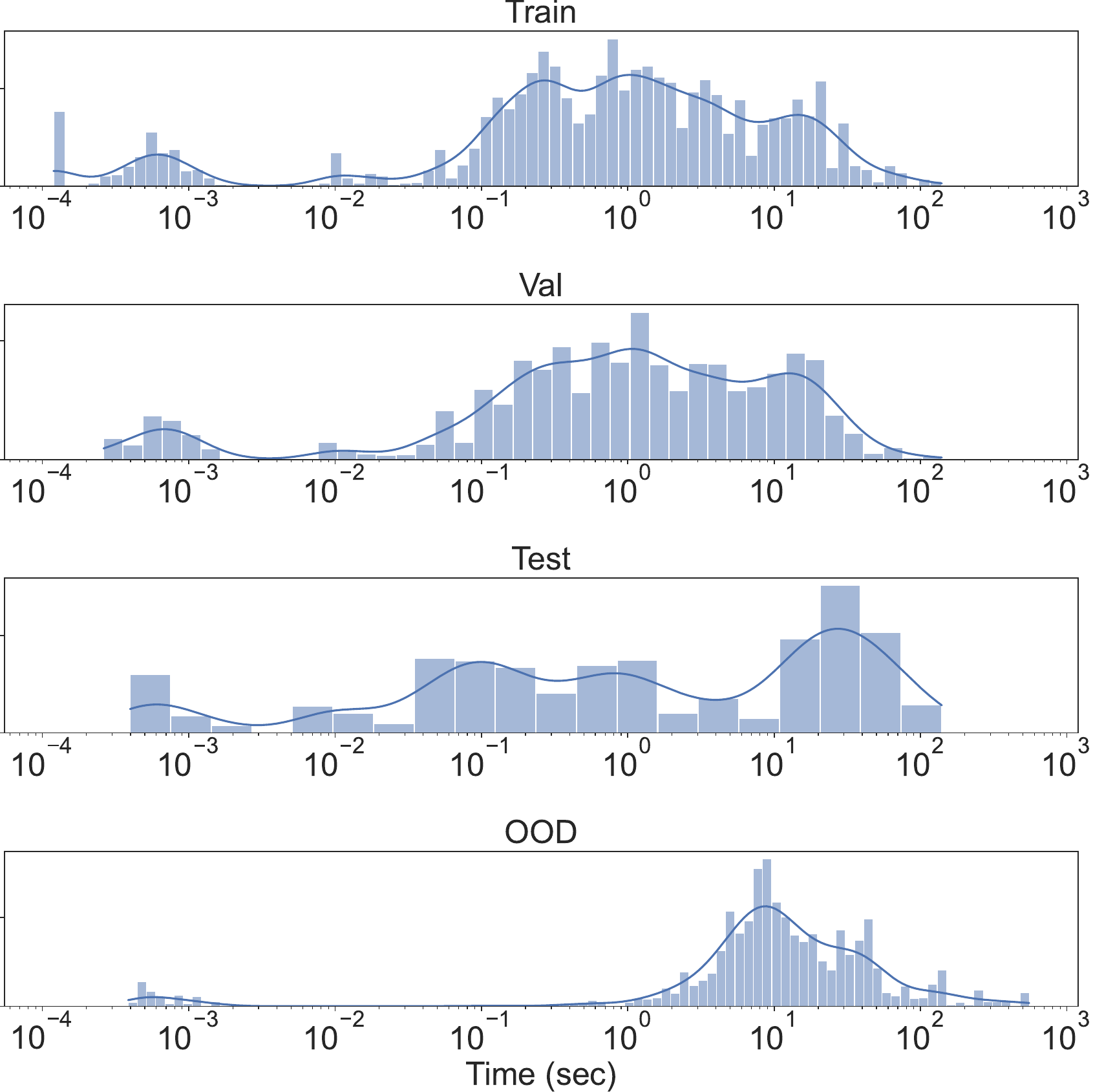}
\caption{Poor generalization on OOD data is, in part, due to the shift in the distribution of plan execution times.}
\label{fig:time_distribution}
\end{figure}

\noindent \textbf{OOD Data.} While analyzing the reasons for significantly large error values on OOD Data (Fig. \ref{fig:generalisation}), we found an interesting observation: the distributions of query latencies on these queries are different (Fig. \ref{fig:time_distribution}). Namely, OOD dataset contained slow queries, the execution times of which were not represented by queries from the Training set. As a result, NN systematically underestimated them. 

\begin{myinsight}
\textbf{Insight (I6)}: \textit{Query latency distribution shift is a potential cause of problematic generalization.
}
\end{myinsight}

\section{Experimental Evaluation of HERO and Comparison with NN-based Models}
\label{chapter:experimental_evaluation}


\textbf{Emulation of NN-based Solutions} As the inference stage in NN-based solutions mirrors the exploration phase of HERO, we augmented them with the parameterized local search procedure from our approach. We implemented a NN architecture that aligns with those in Bao and AutoSteer (see Section \ref{chapter:justification}) which allows for emulating existing hint-based optimizers by properly configuring local search. For example, a NN that explores neighborhoods without semantic-informed operations (i.e., without jointly exploring join and dop hints, or using combinations like INL) serves as an emulation of AutoSteer. A simplified version, in which  where the search is limited to a single iteration, logically emulates QO-Advisor. We will refer to these configurations as \textit{pruned} ones. Bao, on the other hand, implemets a specific case of a local algorithm, where the neighborhood encompasses the entire search space, resulting in \textit{exhaustive} search.

\subsection{Local Search Parameterization}
\label{subchapter:parameterization}

\centeredItalic{What are the optimization capabilities of HERO under different training time budgets?}

\noindent The primary goal of parameterizing the query exploration procedure was to enable balancing between exploration time and the optimality of the discovered hints. To analyze the efficiency of this approach, we tested the performance of HERO on tens of thousands of configurations. For comparison, we developed a metric similar to the $F_{\beta}$ score, which balances the ratio of the achieved boost $c_{boost}$  (relative to the maximum possible boost) to the saved learning time $c_{learning}$:
$$
F_{\beta}(c_{boost}(\text{x}), c_{learning} (\text{x})).
$$

\begin{table*}
\centering
\begin{tabular}{c c c c c c c c c c c}
\toprule
\multicolumn{1}{c}{$\boldsymbol{\beta}$} & 
\multicolumn{1}{c}{\makecell{\textbf{E2E Boost} \\ \textbf{(\% of Opt.) $\uparrow$}}} & 
\multicolumn{1}{c}{\makecell{\textbf{Exploration Time} \\ \textbf{(sec) $\downarrow$}}} & 
\multicolumn{1}{c}{\makecell{\textbf{States} \\ \textbf{(Explored/All)}}} & 
\multicolumn{1}{c}{\makecell{\textbf{Plans} \\ \textbf{(Explored/All)}}} & 
\multicolumn{1}{c}{\textbf{Join}} & 
\multicolumn{1}{c}{\textbf{Scan}} & 
\multicolumn{1}{c}{\textbf{dop}} & 
\multicolumn{1}{c}{\makecell{\textbf{Join $\times$ dop}}} & 
\multicolumn{1}{c}{\textbf{INL}} & 
\multicolumn{1}{c}{\textbf{Iter}} \\
\midrule
0.1  & 40.8  & 894.8  & 339/43392  & 196/7692  & \ding{55} & \ding{55} & \checkmark & \ding{55} & \ding{55} & 1 \\
0.2  & \underline{\textbf{73.1}}  & \underline{\textbf{1417.5}} & 563/43392  & 331/7692  & \ding{55} & \ding{55} & \checkmark & \ding{55} & \checkmark & 2 \\
1.0  & 90.7  & 2912.8 & 1695/43392 & 634/7692  & \checkmark & \ding{55} & \checkmark & \checkmark & \checkmark & 1 \\
2.0  & 94.7  & 3778.3 & 2598/43392 & 896/7692  & \checkmark & \ding{55} & \checkmark & \checkmark & \checkmark & 2 \\
10.0 & \underline{\underline{\textbf{98.5}}}  & \underline{\underline{\textbf{7804.2}}} & 7875/43392 & 2045/7692 & \checkmark & \checkmark & \checkmark & \checkmark & \checkmark & $\infty$ \\
\midrule
Greedy & 78.1 & 9157.7 & -- & -- & \checkmark & \checkmark & \checkmark & \ding{55} & \ding{55} & $\infty$ \\
\bottomrule
\end{tabular}
\caption{
The most effective parameter sets for the Local Search procedure on the JOB benchmark, given various preferences ($\beta$) between performance and training time, highlight that the most critical parameters are those related to parallelism and join selection, followed by the iteration constraint. The least significant parameters are those that control the scan operators. Compared to the greedy algorithm (last row), the Local Search procedure can: a) achieve \underline{similar} performance with a speedup of \underline{approximately six times}, and b) attain \underline{\underline{optimal}} performance within \underline{\underline{the same exploration time}}.
}
\label{tab:boost_performance}
\end{table*}

\noindent Best hint sets for different values of $\beta$ on the JOB benchmark are summarized in Table \ref{tab:boost_performance}. These results confirm our assumptions and provide clear insights:
\begin{itemize}
    \item First, with a small query exploration budget ($\beta=0.1$), the optimal strategy is to adjust the \textit{dop} parameter, proving that the addition of dop-hints into the search space is a sound decision \textbf{(A2)}. 
    \item Second, as the budget increases ($\beta=0.2$), the most efficient strategy is to control \texttt{INL} operator confirming that its misuse is the common mistake of the optimizer \textbf{(A4)}.
    \item Third, with a bigger budget ($\beta=1.0$), the best strategy is to explore all combinations of join types and degrees of parallelism \textit{simultaneously}, ensuring that \textit{all possible pairs} of these parameters are considered at each step. Importantly, the algorithm often stops after the first iteration, thus saving resources \textbf{(A3)}.
    \item Fourth, at even higher budgets ($\beta=2.0$), considering two iterations becomes the most efficient approach, with scan operator hints only being introduced in the final stages ($\beta=10.0$) \textbf{(A1)}. We refer to the final configuration as Local in the remainder of this section.
\end{itemize}   
Additionally, a notable optimization was achieved by avoiding repeated execution of identical plans. Since hints only affect plan construction \textbf{(O2)}, there is no need to explore the same plan twice, even if different hints are used to produce it.

\noindent \textbf{Note on universality.} The presented parameter combinations \textit{are not universally optimal}, as they depend on the specific workload. For example, results from the SQ and TPCH benchmarks, shown in Tables \ref{tab:boost_performance_tpch} and \ref{tab:boost_performance_sq}, reveal that with a very small optimization budget, prohibiting a single join type is optimal for TPCH, while managing INL is key for SQ. This differs from the JOB results. As the budget increases, a general trend emerges across all benchmarks to simultaneously tune join hints and dop (Join $\times$ dop), with scan hints used last. While the greedy algorithm on SQ and TPCH already yields near-optimal results (last rows in the tables), proper tuning of the local search can achieve the same or better performance \textit{several times faster}.

\begin{table}
\centering
\resizebox{\columnwidth}{!}{
\begin{tabular}{c c c c c c c c c c}
\toprule
\multicolumn{1}{c}{$\boldsymbol{\beta}$} & 
\multicolumn{1}{c}{\makecell{\textbf{Boost} \\ \textbf{(\% of Opt.) $\uparrow$}}} & 
\multicolumn{1}{c}{\makecell{\textbf{Exploration} \\ \textbf{(sec)} $\downarrow$}} & 
\multicolumn{1}{c}{\textbf{Join}} & 
\multicolumn{1}{c}{\textbf{Scan}} & 
\multicolumn{1}{c}{\textbf{dop}} & 
\multicolumn{1}{c}{\textbf{Join $\times$ dop}} & 
\multicolumn{1}{c}{\textbf{INL}} & 
\multicolumn{1}{c}{\textbf{Iter}} \\
\midrule
0.1  & 64.5  & 843.7  & \checkmark & \ding{55} & \ding{55} & \ding{55} & \ding{55} & 1 \\
1.0  & 98.7  & 2170.6 & \checkmark & \checkmark & \checkmark & \checkmark & \ding{55} & 1 \\
10.0 & \underline{\textbf{100.0}} & \underline{\textbf{3013.6}} & \checkmark & \checkmark & \checkmark & \checkmark & \ding{55} & \underline{\textbf{2}} \\
\midrule
Greedy & 99.3 & 9986.1 & \checkmark & \checkmark & \checkmark & \ding{55} & \ding{55} & $\infty$ \\
\bottomrule
\end{tabular}
}
\caption{Best parameters for the Local Search under varying training budgets on the TPCH benchmark show that just \underline{\textbf{two}} hints—related to joins and dop—are sufficient to achieve \underline{\textbf{maximum}} performance, with \underline{three times} the exploration speed of the greedy algorithm.}
\label{tab:boost_performance_tpch}
\end{table}

\begin{table}
\centering
\resizebox{\columnwidth}{!}{
\begin{tabular}{c c c c c c c c c}
\toprule
\multicolumn{1}{c}{$\boldsymbol{\beta}$} & 
\multicolumn{1}{c}{\makecell{\textbf{Boost} \\ \textbf{(\% of Opt.) $\uparrow$}}} & 
\multicolumn{1}{c}{\makecell{\textbf{Exploration} \\ \textbf{(sec)} $\downarrow$}} & 
\multicolumn{1}{c}{\textbf{Join}} & 
\multicolumn{1}{c}{\textbf{Scan}} & 
\multicolumn{1}{c}{\textbf{dop}} & 
\multicolumn{1}{c}{\textbf{Join $\times$ dop}} & 
\multicolumn{1}{c}{\textbf{INL}} & 
\multicolumn{1}{c}{\textbf{Iter}} \\
\midrule
0.1  & 40.5  & 1308.8 & \ding{55} & \ding{55} & \ding{55} & \ding{55} & \checkmark & 1 \\
0.2  & 59.3  & 1727.3 & \checkmark & \ding{55} & \ding{55} & \ding{55} & \ding{55} & 1 \\
1.0  & 80.5  & 2218.1 & \checkmark & \ding{55} & \ding{55} & \ding{55} & \ding{55} & $\infty$ \\
2.0  & 92.4  & \underline{\textbf{5222.1}} & \checkmark & \ding{55} & \checkmark & \checkmark & \checkmark & 2 \\
10.0 & 94.1  & 9866.8 & \checkmark & \checkmark & \checkmark & \checkmark & \checkmark & $\infty$ \\
\midrule
Greedy & 88.2 & 9157.8 & \checkmark & \checkmark & \checkmark & \ding{55} & \ding{55} & $\infty$ \\
\bottomrule
\end{tabular}
}
\caption{Best parameters for the Local Search under different training budgets on the SQ benchmark. As shown, with the right parameters, it can achieve the performance of the greedy algorithm in about \underline{half the time}.}
\label{tab:boost_performance_sq}
\end{table}

\begin{myconclusion} 
    \noindent \textbf{Summary.}  \textit{The parameterized Local Search procedure of HERO efficiently manages the trade-off between exploration and performance, achieving nearly optimal results across all three benchmarks. The procedure adapts to various budgets, minimizing redundant exploration while ensuring consistent performance. It is important to note that there is \underline{no universally} optimal configuration for local search, so it should be tailored to the specific characteristics of the workload.}
\end{myconclusion}

\subsection{Static Workload}

\centeredItalic{In ideal conditions, can NN-based solutions outperform HERO?}

\label{subchapter:static_workload}

\noindent \textbf{Offline Scenario.}  The results of comparing HERO and NN-based solutions in conditions, where all data was observable at training and the exploration budget was unlimited, are presented in Table \ref{tab:ideal_static_workload}. Notably, even in these ideal conditions, NN-based solutions experienced timeouts and led to query latency degradations. The primary advantage of HERO over NN-based approaches lies in its inference which \textit{is more than five times faster}.

\begin{myconclusion}
    \noindent \textbf{Summary.} \textit{Even in ideal conditions, NN-based solutions lead to degradations, while HERO makes reliable hint recommendations, while its main performance advantage is in utilizing query-specific information to reduce the inference time.}
\end{myconclusion}

\begin{table*}
\centering
\begin{tabular}{c c c c c c c c}
\toprule
\textbf{Model} & \makecell{\textbf{Inference} \\ \textbf{Settings}} & \makecell{\textbf{E2E Boost} \\ \textbf{(\%) $\uparrow$}} & \makecell{\textbf{E2E Boost} \\ \textbf{(\% of Opt.) $\uparrow$}} & \makecell{\textbf{Ex Boost} \\ \textbf{(\% of Opt.) $\uparrow$}} & \makecell{\textbf{Timeouts} \\ \textbf{(\%) $\downarrow$}} & \makecell{\textbf{Degradations} \\ \textbf{(\%) $\downarrow$}} & \makecell{\textbf{Inference} \\ \textbf{(sec)} $\downarrow$} \\
\midrule
HERO  & Local          & 64.6  & 98.4  & 98.9  & 0.0 & 0.0 & 26.7 \\
NN    & Pruned Local   & 56.2  & 85.7  & 94.1  & 0.9 & 2.7 & \underline{\textbf{58.5}} \\
NN    & Local         & 37.5  & 57.2  & 97.5  & \textcolor{Maroon}{\textbf{4.4}} & \textcolor{Maroon}{\textbf{5.3}} & \textcolor{Maroon}{\textbf{170.7}} \\
NN    & Pruned Greedy  & 28.7  & 43.7  & 47.0  & 0.0 & 1.8 & \underline{\textbf{71.5}} \\
NN    & Greedy         & 20.2  & 30.8  & \underline{\underline{\textbf{68.4}}}  & 1.8 & 2.7 & \textcolor{Maroon}{\textbf{180.4}} \\
NN    & Exhaustive     & -17.8 & -27.0 & 98.3  & \textcolor{Maroon}{\textbf{8.8}} & \textcolor{Maroon}{\textbf{10.6}} & \textcolor{Maroon}{\textbf{475.2}} \\
\bottomrule
\end{tabular}
\caption{Comparison of HERO and NN-based models with different Local Search algorithm parameters for inference on the JOB benchmark. All models were trained under \textit{ideal conditions}: static workloads and unlimited training budget. HERO achieves a higher performance because of faster inference while also being more reliable. NN-based approaches, however, suffer from \textcolor{Maroon}{\textbf{risks of degradation}} and \textcolor{Maroon}{\textbf{slower inference}}. Nevertheless, using the HERO-inspired \underline{pruned} version of the Local Search algorithm helps to mitigate these issues by significantly improving the inference speed and reducing the likelihood of degradation. The table also shows that the greedy algorithm \underline{\underline{does not provide}} near-optimal hint sets.}
\label{tab:ideal_static_workload}
\end{table*}

\begin{figure}
\centering
\includegraphics[width=\columnwidth]{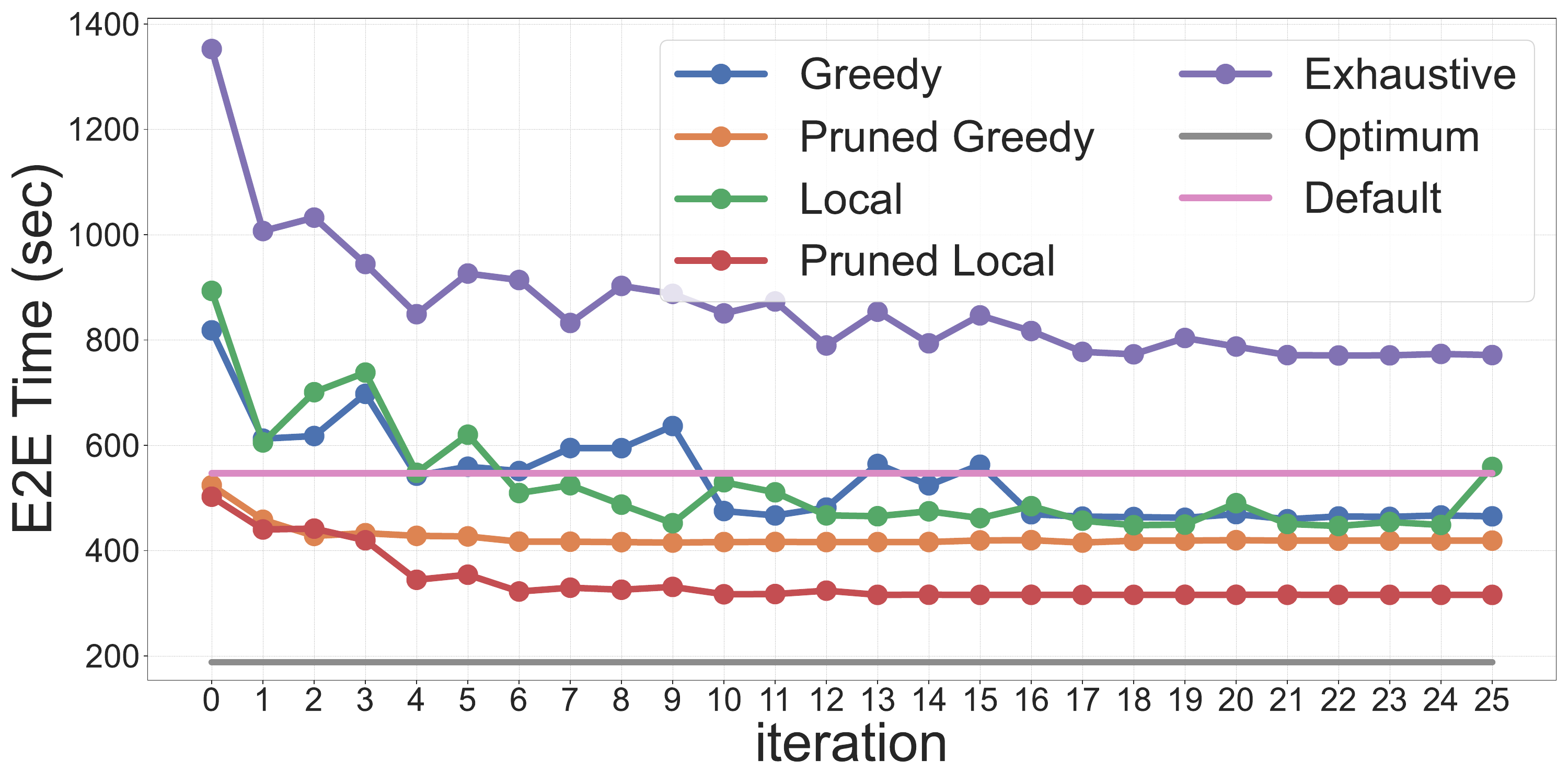}
\caption{Results of online scenario simulation on JOB benchmark. Before the simulation, we used \textit{pretraining} on default plans. Note that  only the pruned versions achieve workload speed up after after this pretraining.}
\label{fig:all_dops_online_job}
\end{figure}

\noindent \textbf{Online Scenario.} We simulated an online learning scenario for NN-based algorithms in which we iterated through all JOB queries, selecting hints based on the current prediction of NN and updating its weights with the accumulated experience. The results presented in Figure \ref{fig:all_dops_online_job} highlight three key findings:

\begin{itemize}
    \item As the search space grows, the inference overhead for Bao (Exhaustive) and AutoSteer (Greedy) becomes excessively large, making the approaches impractical, while the workload speedup obtained by QO-Advisor (Pruned Greedy) is minor.
    \item The pruned version of our proposed Local Search algorithm alleviates this issue, significantly reducing inference overhead and improving NN-based methods.
    \item The overall improvement remains well below what was achieved in the offline mode (Tab. \ref{tab:ideal_static_workload}). We believe that the reason is that in online learning, the model influences the data it later trains on, leading to local optima and the underestimation of higher-performing states.
\end{itemize}  

\begin{myconclusion}
    \noindent \textbf{Summary.} \textit{While NN-based models support online lear\-ning, the high inference overhead \textbf{(C1)} and the insufficient representativeness of the collected data \textbf{(C2)} make existing approaches impractical. Only the pruned versions show an acceptable performance, but it is well below what is achieved in the offline training mode.}
\end{myconclusion}

\subsection{Dynamic Workload}
\label{subchapter:dynamic_workload}

\centeredItalic{Can HERO generalize to new queries?}

\begin{table*}
\centering
\begin{tabular}{c c  c c c  c c c  c c c  c c c}
\toprule
& & \multicolumn{3}{c}{Structure Split} & \multicolumn{3}{c}{Fast-to-slow split} & \multicolumn{3}{c}{Random Split} & \multicolumn{3}{c}{Slow-to-fast split}\\
\cmidrule(lr){3-5} \cmidrule(lr){6-8} \cmidrule(lr){9-11} \cmidrule(lr){12-14}

\textbf{Model} & \makecell{\textbf{Inference} \\ \textbf{Settings}} & 

\makecell{\textbf{E2E} \\ \textbf{(\%) $\uparrow$}} & \makecell{\textbf{T/Os} \\ 
\textbf{(\%) $\downarrow$}} & \makecell{\textbf{Degr.} \\ \textbf{(\%) $\downarrow$}} 

& \makecell{\textbf{E2E} \\ \textbf{(\%) $\uparrow$}} & \makecell{\textbf{T/Os} \\ 
\textbf{(\%) $\downarrow$}} & \makecell{\textbf{Degr.} \\ \textbf{(\%) $\downarrow$}}

& \makecell{\textbf{E2E} \\ \textbf{(\%) $\uparrow$}} & \makecell{\textbf{T/Os} \\ 
\textbf{(\%) $\downarrow$}} & \makecell{\textbf{Degr.} \\ \textbf{(\%) $\downarrow$}} 

& \makecell{\textbf{E2E} \\ \textbf{(\%) $\uparrow$}} & \makecell{\textbf{T/Os} \\ 
\textbf{(\%) $\downarrow$}} & \makecell{\textbf{Degr.} \\ \textbf{(\%) $\downarrow$}}\\
\midrule


NN    & Pruned Local   
& \underline{\textbf{31.7}}  & \textcolor{Maroon}{\textbf{12.1}} & \textcolor{Maroon}{\textbf{19.2}}
& 21.5 & 0.0  & 9.1
& 15.7  & \textcolor{Maroon}{\textbf{10.2}} & \textcolor{Maroon}{\textbf{23.9}}
& -1.8   & 9.1  & \textcolor{Maroon}{\textbf{21.2}}\\

NN    & Pruned Greedy  
& 18.8  & 0.8  & 8.8
&  19.4  & 0.0  & 6.1
& 13.8  & 1.1  & \textcolor{Maroon}{\textbf{17.0}}
& 0.5  & 0.0  & \textcolor{Maroon}{\textbf{21.2}}\\

HERO  & Local
& 15.8  & \underline{\underline{\textbf{4.2}}}  & \underline{\underline{\textbf{4.2}}}
& 0.0   & 0.0  & 0.0
& 4.8   & 0.0  & 0.0
& 0.0   & 0.0  & 0.0\\

NN    & Local          
& 12.5  & \textcolor{Maroon}{\textbf{24.6}} & \textcolor{Maroon}{\textbf{27.5}} 
&  -4.8  & \textcolor{Maroon}{\textbf{18.2}} & \textcolor{Maroon}{\textbf{24.2}}
& \textcolor{Maroon}{\textbf{-17.9}} & \textcolor{Maroon}{\textbf{24.7}} & \textcolor{Maroon}{\textbf{31.4}}
& 15.7  & \textcolor{Maroon}{\textbf{10.2}} & \textcolor{Maroon}{\textbf{23.9}}\\

NN    & Greedy         
& 16.6  & \textcolor{Maroon}{\textbf{11.7}} & \textcolor{Maroon}{\textbf{17.5}}
& 9.3   & 3.0  & \textcolor{Maroon}{\textbf{12.1}}
& \textcolor{Maroon}{\textbf{-20.7}} & \textcolor{Maroon}{\textbf{12.6}} & \textcolor{Maroon}{\textbf{23.0}}
& \textcolor{Maroon}{\textbf{-93.9}} & 9.1  & \textcolor{Maroon}{\textbf{24.2}}\\

NN    & Exhaustive     
& \textcolor{Maroon}{\textbf{-10.7}} & \textcolor{Maroon}{\textbf{34.6}} & \textcolor{Maroon}{\textbf{35.8}}
& \textcolor{Maroon}{\textbf{-57.0}} & \textcolor{Maroon}{\textbf{45.5}} & \textcolor{Maroon}{\textbf{45.5}}
& \textcolor{Maroon}{\textbf{-89.0}} & \textcolor{Maroon}{\textbf{30.5}} & \textcolor{Maroon}{\textbf{35.4}}
& \textcolor{Maroon}{\textbf{-234.8}} & \textcolor{Maroon}{\textbf{21.2}} & \textcolor{Maroon}{\textbf{24.2}}\\
\bottomrule
\end{tabular}
\caption{Metrics on dynamic workloads with different splitting strategies. In particular cases NN provides a \underline{\textbf{higher}} performance which comes at the price of a higher degradation \textcolor{Maroon}{\textbf{risk}} for other queries. HERO, on the other hand, may occasionally arrive at hint sets that \underline{\underline{slow down queries}}, but its transparency, as we previously described in \S \ref{subchapter:removing_nn}, allows us to simply turn off the predictions in regions where they negatively impact on workload.}
\label{tab:dynamic_workload}
\end{table*}

\noindent 
We also conducted experiments where the model performance was measured on \textit{unseen} queries. Based on our observations about the properties of data that help or hinder knowledge generalization, we considered several ways of splitting the data: \textbf{(a)} splitting randomly, \textbf{(b)} splitting queries within a group with the same default plan structure, and \textbf{(c)} splitting queries by the execution time of their default plan. Results are given in Table \ref{tab:dynamic_workload}. Three important observations can be made from them:
\begin{itemize}
    \item The versions with pruning consistently achieved the highest performance boosts. Not only did this reduce inference time, but it also significantly limited the search space, \textit{reducing the likelihood of selection of bad plans}.
    \item The presence of queries with similar plan structures in the training set greatly increases overall gains, supporting our insight \textbf{(I5)} on the importance of plan structure for generalization ability. NN-based solutions can deliver higher performance boosts but often come with risks of degradations and timeouts. In contrast, HERO offers more moderate, but safer improvements.
    \item Generalization from fast queries to slower ones was significantly more successful than the opposite. This is likely because exploration requires finding plans with shorter execution times, which are more likely to be present by the training set if it contains fast queries, and this is partly explained by our insight \textbf{(I6)}.
\end{itemize}  

\begin{myconclusion}
    \noindent \textbf{Summary.} \textit{For dynamic workloads, HERO offers a more reliable and stable solution. In contrast, existing NN-based solutions face risks of degradations and timeouts, which makes their practical application challenging. However, in particular cases these models can achieve significantly higher performance boosts. Using Local Search techniques, especially with pruning, helps to reduce the inference time and enhances the reliability of NN-based predictions.}
\end{myconclusion}

\section{Conclusion}
\label{chapter:conclusions}

In this paper, we proposed HERO, an efficient and reliable hint-based query optimizer which sets itself apart from existing solutions by  \textbf{(a)} replacing NN with a more interpretable, reliable, and controllable ensemble of con\-text-aware models, and by \textbf{(b)} introducing a new parameterized local search procedure that efficiently explores queries and supports adjusting balance between training time and performance gains. We presented detailed experimental results showing that HERO sets a new level for query optimization with coarse-grained hints. 

We have also provided an in-depth analysis of architecture, benefits, and limitations of the existing NN-based approaches to hint-based optimization and showed that several design decisions from HERO can be used to improve them. We believe that the results of our study will facilitate next steps to unlocking the full potential of hint-based learned query optimization. 

Our work has primarily focused on coarse-grained hints, which affect the entire query plan, but may not achieve optimal query acceleration. Fine-grained hint mechanisms already exist (e.g., pg\_\-hint\_\-plan \cite{pghintplan} for PostgreSQL), which allow for setting the range of admissible operators at plan nodes and even taking control over the cardinality model and join order. However, with fine-grained hints, the issue of search space explosion becomes even more pronounced -- the dimensionality increases not only with the size of the hint set but also with the complexity of the query itself. Nevertheless, we believe that fine-grained hints have a big potential, and future work could leverage this approach to further improve query optimization.


\bibliographystyle{ACM-Reference-Format}
\bibliography{references}

\end{document}